\documentclass[12pt,aps,amsmath,latexsym,amsfonts]{JHEP3}

\usepackage{bm}
\usepackage{graphicx}
\usepackage{amssymb}
\usepackage{amsmath}

\newcommand{\be}{\begin{equation}}
\newcommand{\ee}{\end{equation}}
\newcommand{\bea}{\begin{eqnarray}}
\newcommand{\eea}{\end{eqnarray}}

\def\real{\mathbb R}

\title{Cosmic superstring trajectories in warped compactifications}
\author{
Anastasios Avgoustidis$^{1}$\thanks{Email: 
a.avgoustidis@damtp.cam.ac.uk}\ ,
Sarah Chadburn$^{2}$\thanks{Email: s.e.chadburn@durham.ac.uk}\ , and
Ruth Gregory$^{2,3}$\thanks{Email: r.a.w.gregory@durham.ac.uk}\\
$^1$School of Physics and Astronomy, University of Nottingham, \\
University Park, 
Nottingham NG7 2RD, UK\\
$^2 $Centre for Particle Theory, Durham University,\\ 
South Road, Durham, DH1 3LE, UK\\
$^4 ${\it Perimeter Institute, 31 Caroline St, Waterloo, Ontario N2L 2Y5,
Canada}
}

\abstract{
We explore the generic motion of cosmic (super)strings when the 
internal compact dimensions are warped, using the Klebanov-Strassler
solution as a prototypical throat geometry. We find that there is 
no dynamical mechanism which localises the string at the tip 
of the throat, but rather that the motion seems to explore both internal 
and external degrees of freedom democratically. This indicates that cosmic 
(super)strings formed by inflationary brane-antibrane annihilation
will have sufficient internal motion for the gravitational wave 
signals from the string network to be suppressed relative to the 
signal from a `standard' cosmic string network. 
}

\preprint{DCPT-12/19}

\begin{document}

\section{Introduction}

Cosmic strings, \cite{CS}, have had a checkered past in terms 
of their relevance or popularity in particle cosmology. 
Initially, these strings were relic topological defects of some 
Grand Unified phase transition in the early universe, \cite{ZelKib}, and 
proposed as an alternative to inflation. However, the microwave background 
perturbations show that strings cannot be responsible for structure
formation, \cite{CMBrout}, although can nonetheless be present as
a sub-dominant component, \cite{subdom}. 
Recently, cosmic strings have had a renaissance
as by-products of string theoretic brane inflationary scenarios 
\cite{braneinf,KKLMMT,reviews,BCS,JST}.
The basic picture here is that inflation is driven by a mobile
D-brane in a warped extra dimension 
scenario. Inflation
ends (typically) by brane/anti-brane annihilation, a by-product of
which are cosmic superstrings, \cite{JST}. Although, when formed, 
these objects are inherently string theoretic, and can have interesting
and distinctive properties, it is assumed that they persist to low energies
as topological defects somewhat similar to the more traditional
cosmic strings. As such, if detected, they could provide indirect 
evidence of string theory \cite{CPV, Hind_Rev}.

The basic cosmological picture is that cosmic strings form 
a network of loops and long strings, whose evolution is determined
by rules for intercommutation \cite{IC}, or how crossing strings interact, 
and an effective action governing the motion of the string, which is
well approximated by the zero width action, \cite{NG,FWC}:
\be\label{NGaction}
S = -\mu \int d^2 \sigma \; \sqrt{\gamma}
\ee
where $\mu$ is the mass per unit length of the string, and $\gamma_{AB}$
the induced metric on the worldsheet.
Incorporating gravitational effects via a linearized approximation
indicates how fast energy is lost from the network, \cite{gwave,gwave2}, 
and putting all these pieces together gives a scaling
behaviour for the string network, \cite{NET}.
Cosmic (super)strings, \cite{CSSrev},
are generally modelled in a similar fashion, but have some 
interesting differences as they are derived from
a higher dimensional theory, string theory, and have additional physics
arising as a result. The most important difference is that these strings need
not intercommute when they apparently intersect in our 4D world, \cite{ICSC}, 
resulting in a denser network, \cite{JST,Sak_lowP,AS2}.
Cosmic superstrings can
also have junctions, \cite{ICSC,CKS,GWCSS,JCT}, 
although the implications for the network have been less well explored. 
For recent progress, both on the numerical and analytic fronts, 
see \cite{zipnetA,zipnetN,zipnetP}.

The particular phenomenon we are interested in in this paper is 
the effect of the extra dimensions on the kinematics of the string
network. The motion of the string can be thought of as left and
right moving waves along the string, \cite{KT2}, thus if there is
internal motion, the projection of these waves onto our
non-compact four-dimensional spacetime leads to a lower 
apparent velocity, \cite{AS}. (This effect is analogous to string 
currents in superconducting strings~\cite{Nielsen,Witten,VilShel}.)
While this does not have so strong an effect on the shape 
of general loops,
the impact on highly relativistic events such as cusps is dramatic.
Cusps are transitory events, where constructive interference between the left 
and right moving modes causes a point of the string instantaneously to reach 
the speed of light, however, if the strings are moving slower, they will
not only fail to reach the speed of light instantaneously in the noncompact
dimensions, but in fact also have a reduced probability of such a
relativistic event occurring. 
Another singular feature of cosmic string motion are kinks, or sharp
`corners' on the string representing a discontinuity in the wave velocity.
They are formed as relics of intercommutation, and 
because they are a feature of only one wavefront,
they persist and move along the string, \cite{ACKinks}. 

Since both kinks and cusps represent
a certain level of singularity of the worldsheet swept out by
the string, when including gravitational backreaction they
generate strong and distinct gravitational wave signals, 
\cite{gwave,gwave2,FAN,GWL,Eloss,DV}, which have been used to place
bounds on the tension of the string and the inflationary scenario,
\cite{obsv}.
Recently, \cite{CCGGZ,CCGGZ2}, the implication of
the motion in the extra dimensions was explored by modelling
the extra dimensions as flat, and considering the impact of the
reduced wave velocity in the non-compact directions. 
The cusp signal was found to be significantly damped, primarily due to the
fact that motion in higher dimensions has a significantly reduced
probability of cusp formation. The kink signal was also
reduced, though less dramatically. In each case, a reduction in
signal can be linked to the presence of internal velocities, thus
any possible mechanism which would damp or dry out motion in the
internal dimensions would significantly alter our previous
conclusions.

In this paper, we consider the impact of warped internal 
dimensions on the motion of the cosmic superstrings. 
It is commonly asserted that the warping 
of spacetime causes cosmic strings to be localised in 
the internal dimensions; that is, at late times, their 
positions can be expected to get fixed 
at the tip of the warped throat, with no freedom to 
move on the internal manifold.  This can be understood
by noting that the string positions in the internal dimensions 
are worldsheet scalar fields~\cite{Nielsen}, which have 
a massive potential in the presence of warping.  At low energies
therefore, one might expect that such additional fields would 
acquire masses and stabilise, presumably corresponding to the
string lying at the bottom of the warped throat. However, it is
not entirely clear that the string can be dynamically stabilised
at the potential bottom in the absence of a strong damping mechanism 
for internal excitations~\cite{tasos2008}. Moreover,
if the string is displaying characteristics derived from its
higher dimensional inheritance, such as a low intercommutation 
probability, then it seems entirely possible that it will also
display other consequences of its higher dimensional nature, such
as these internal degrees of freedom. In the absence of an explicit
concrete construction of the low energy degrees of freedom,
it seems reasonable to explore
the consequence of internal motion for the cosmic string.

Here, we examine this issue in more detail by studying the classical 
motion of strings in warped spacetimes. This is motivated by the fact 
that, in many realistic compactifications, the relevant compactification 
scales can be much larger than the string scale, which sets the scale
of the string thickness. Thus, at late times, when 
the strings are part of a cosmological network and are effectively
topological defects existing at low energies, it seems reasonable to
assume that if they are semi-classical objects as far as our 4D universe 
is concerned, then they also have a semi-classical nature in the 
extra dimensions. We therefore allow for the string to be free to move 
in the internal dimensions, and study the dynamics of 
string loops in a warped throat modelled by the Klebanov-Strassler (KS) 
solution \cite{KS}.  We find that, although there is an attractive 
force towards the bottom of the throat, there is no friction to 
ensure stabilisation.  On the contrary, there is continuous exchange 
of energy between the external and internal sectors, reinforcing motion 
in the internal manifold.  This provides strong evidence that string loops can 
have significant motion in the internal dimensions, even in a strongly 
warped spacetime. 

\section{Strings on a warped internal manifold} 
\label{modelling}

In order to explore the effect of warped internal dimensions on the motion
of the string, we use the concrete example of the warped deformed conifold, 
or Klebanov-Strassler, solution \cite{KS}. This is an exact supergravity
solution with D3 and wrapped D5 branes, which interpolates from a 
regular $\real^3 \times S^3$ tip, to an $\real \times T^{1,1}$ cone in the UV. 
The ten dimensional metric is: 
\be
\label{10Dmetric}
ds_{10}^2\equiv G_{ab}dx^a dx^b = h^{-1/2}{g_{\mu\nu}dx^{\mu}dx^{\nu}}
- h^{1/2}{{\tilde g}_{mn}dy^{m}dy^{n}},
\ee
where $g_{\mu\nu}(x)$ is the 4D spacetime metric 
and ${\tilde g}_{mn}(y)$ the 6D internal metric, whose explicit form
will be given below. The warp factor, $h$, depends only on the radial 
internal direction $\eta$ (see Eq.~(\ref{KS6}) below), 
and is given by:
\be
\label{KSh}
h = 2 (g_{s}M\alpha^{\prime})^{2}\; \epsilon^{-8/3} \;  I(\eta),
\ee
where
\be
\label{Iofeta}
I(\eta)\equiv \int_{\eta}^{\infty}{dx\frac{x\coth x-1}{\sinh^{2}x}}
(\sinh x \cosh x -x)^{1/3}.
\ee
Here $M$ is a compactification parameter representing the number
of dissolved D5 branes in the background, and $\epsilon$ is a
dimensionful parameter measuring the deformation of the conifold. 
The string coupling and string scale are given as usual by $g_s$ 
and $\alpha'$. 

Our four-dimensional spacetime $g_{\mu\nu}dx^{\mu}dx^{\nu}$
will be taken either to be Minkowski spacetime, as in the
original KS solution, or an FRW universe, when we consider
the effect of cosmological expansion.
The internal manifold is the warped throat, and has the form:
\bea
{\tilde g}_{mn} dy^m dy^n &=&
\frac{\epsilon^{4/3}}{2} K(\eta)\Big[
\frac{1}{3K(\eta)^3}\{d\eta^2+(g^{5})^2\}
+\cosh^2\frac{\eta}{2}\{(g^{3})^2+(g^ {4})^2\}\nonumber \\
&& \hskip 2cm + \sinh^2\frac{\eta}{2}\{(g^{1})^2+(g^{2})^2\}\Big],
\label{KS6}
\eea
where $\eta$ is a radial coordinate in which the other metric functions
have analytic expressions
\be
\label{KSfn}
K(\eta)=\frac{(\sinh\eta\cosh\eta-\eta)^{1/3}}{\sinh\eta}\,.
\ee 
The $g^i$'s are forms representing the angular directions, 
typically expressed by
\be
\label{gdef}
g^{1,3}=\frac{e^{1}\mp e^{3}}{\sqrt{2}},\;\;\;\ g^{2,4}=\frac{e^{2}\mp
e^{4}}{\sqrt{2}},\;\;\;\ g^{5}=e^{5}
\ee
with
\bea                                                                  
\label{edef}
e^{1} & = & -\sin\theta_{1} d\phi_{1},\;\;\;\;\;\;\  e^{2}= d\theta_{1},
\;\;\;\;\;\;\  e^{3}= \cos\psi \sin\theta_{2}d\phi_{2}-\sin\psi d\theta_{2},
\nonumber \\
e^{4} &= & \sin\psi\sin\theta_{2}d\phi_{2}+\cos\psi d\theta_{2}, 
\;\;\;\;\;\;\
e^{5}= d\psi+\cos\theta_{1}d\phi_{1}+\cos\theta_{2}d\phi_{2}.
\eea
(For details on the warped deformed conifold, and coordinate
systems, see e.g.~\cite{KS, coni}.)

In addition, there are NSNS and RR fluxes in the throat, given by
\bea
B_2 &=& g_sM \frac{(\eta\coth\eta-1)}{2\sinh \eta} \left [
(\cosh\eta -1) g^1 \wedge g^2 + (\cosh\eta+1)g^3\wedge g^4 \right] 
\label{Bflux}
\\
F_3 &=& \frac{M}{2} \left [ 2 g^3 \wedge g^4 \wedge g^5 + 
d \left ( (1-\eta\,{\rm csch}\,\eta) [ g^1 \wedge g^3 + g^2 \wedge g^4] \right)
\right ] \,.
\label{Fflux}
\eea 

The exact KS solution is an infinite throat, however, it is presumed to
be a good approximation to a finite throat glued to a compact Calabi-Yau 
internal manifold at some UV scale $r_{UV} \simeq \epsilon^{2/3} 
\int_0^{\eta_{UV}} d\eta /\sqrt{6}K$. This scale sets the UV/IR 
hierarchy of the throat
$h_0/h_{UV}\sim 0.302\, e^{\textstyle{\frac43}\eta_{UV}}/\eta_{UV}$,
and the size of the internal manifold. This latter relation provides
a lower bound on the four dimensional Planck scale in terms
of the string scale and compactification parameters
\be
M_p^2 \gtrsim \frac{\epsilon^{4/3}M^2}{3(2\pi)^4\alpha^{\prime2}}
\int_0^{\eta_{UV}} I(\eta) \sinh^2\eta d\eta
\label{BMBD}
\ee
obtained by performing a dimensional reduction and integrating 
out the Einstein action over the internal manifold, \cite{BM}.

It is worth pausing to emphasize the geometrical consequences of 
this throat geometry, as these feed into the string equations of
motion. The prefactor $h$ is very sharply peaked 
at the base of the throat, and thus leads to a large hierarchy
between our four-dimensional noncompact physics and the underlying
ten-dimensional scales. For example, if we consider a length
scale, such as the string effective width $w_4$ in our universe, 
then near the tip this is equivalent to an
internal range of $\eta$ given by $w_4^2 \sim (g_s M \alpha')^2
\epsilon^{-4/3} (\delta \eta)^2$, or 
\be
w_4^2 M_p^2 \gtrsim \frac{\alpha' g_s^2M^4}{3(2\pi)^4}
\epsilon^{-4/3}  (\delta \eta)^2 
\label{ltoeta}
\ee
(using (\ref{BMBD}), and replacing the integral with an estimate in terms
of $r_{UV}^2 = {\cal O}(\alpha')$). For the string width scale, we expect
$w_4^2 M_p^2 \simeq 1/(G\mu) \sim 10^7 - 10^{12}$, 
and for typical compactification data considered in brane
inflation models (see \cite{GK} for a translation of this
to the KS parameters used here) the prefactor on the RHS of
(\ref{ltoeta}) is $10^{12}-10^{14}$, hence $(\delta \eta)^2$
is likely to be extremely small.

To get the equations of motion, we therefore take the zero width
classical string effective action, \cite{NG,FWC}, i.e.\
the Nambu action (\ref{NGaction}).
Because our metric is nontrivial, we cannot take the usual temporal 
conformal gauge used by Kibble and Turok, \cite{KT2}, in which the 
induced metric on the string worldsheet is conformally flat with 
worldsheet and bulk time identified, and in which the motion of the
string reduces to left and right moving uncoupled waves. Instead, we
choose the transverse temporal gauge:
$\sigma^0 = X^0\equiv t$, $\dot{X}^{a}X'_{a} = 0$. This identifies 
worldsheet time with background time and imposes diagonality on the
worldsheet metric $\gamma_{AB} = X^a_{,A} X^b_{,B} G_{ab}$.
Variation of the Nambu action gives a wave equation for $X^a$:
\be
\begin{aligned}
\Box X^a + \Gamma^a_{bc} X^b_{,B} X^c_{,C} \gamma^{BC} = 0 \ \ &\Rightarrow& \; \\
\frac{\partial}{\partial t}\left(\frac{\dot{X}^{a}
X'^{2}}{\sqrt{-\gamma}}\right) 
+ \frac{\partial}{\partial \sigma}\left(\frac{X'^{a}\dot{X}^{2}}
{\sqrt{-\gamma}}\right)
+ \frac{1}{\sqrt{-\gamma}}\Gamma^{a}_{bc}\left(
X'^{2}\dot{X}^{b}\dot{X}^{c} + \dot{X}^{2}X'^{b}X'^{c}
\right)    &=&  0 \,,
\end{aligned}
\label{generalEoM}
\ee
where a dot denotes differentiation with respect to time and prime 
the spacelike worldsheet coordinate $\sigma^1\equiv \sigma$. 

It is now straightforward to see why the conformal gauge cannot
be simultaneously chosen with the synchronous gauge, even if our
four-dimensional universe is flat, by examining the $a=0$ equation
which reduces to
\be
\frac{\partial}{\partial t}
\left ( \sqrt{\frac{-X'^{2}}{h\dot{X}^2}} \right ) =0
\label{t-eom}
\ee
for the case $g_{\mu\nu} = \eta_{\mu\nu}$. Clearly, if $h$ varies
significantly over the timescales of interest, then the synchronous
gauge will not be a good approximation. 

In order to explore the range of string motion, we first look at
some simple trajectories in a background where the 
non-compact dimensions are flat, before considering
how to extract more general trajectories and to include more
physics such as cosmological expansion and gravitational backreaction.

\section{String motion in a static spacetime} \label{static}

We first consider strings in a Minkowski background, $g_{\mu\nu} = 
\eta_{\mu\nu}$ in order to derive the basic behaviour of a loop.
If the full spacetime is flat, then the Kibble-Turok, \cite{KT2},
method for building general solutions can be applied. In the
warped background however, this is no longer the case, and
we must look carefully at how the internal and external directions
interact through the warp factor. It is useful to first consider a 
simple trajectory to explore the main issues associated with warping, 
thus we restrict the internal motion to only two directions, 
looking at movement in the radial $\eta$-direction and one angular direction. 
In order that the angular motion is a stable trajectory (i.e.\ that it
does not induce motion in any of the other angular directions) we must
take it to be along a great circle within the conifold, and in particular
avoiding any coordinate singularities, such as polar singularities. 
We explicitly choose
$\theta_1 = \theta_2 = \frac{\pi}{2}$, $\psi = \pi$, $\phi_1 = \phi_2 
= \phi$, thus having the string move around the non-contractible
$S^3$, which is a consistent angular trajectory.
The metric effectively becomes:
\be 
ds^2 = h^{-1/2}\left(dt^2 - d\textbf{x}^2\right) 
- h^{1/2}\epsilon^{\frac{4}{3}}\left[\frac{d\eta^2}{6K^2
(\eta)}+B(\eta)d\phi^2\right]\,, 
\ee
where $B(\eta) = K(\eta)\cosh^2\left(\frac{\eta}{2}\right)$. 
By modelling the motion of string loops in this 
spacetime, we hope to capture essential aspects of their dynamics 
in the full 10D spacetime. The string 
worldsheet can be written as:
\be
X^a(t,\sigma) = (t, \textbf{x}(t,\sigma), \eta(t,\sigma), \phi(t,\sigma)) \,.
\label{twod}
\ee
Substituting this into the equation of motion, (\ref{generalEoM}), gives 
the explicit set of equations for the
spacetime coordinates of the string worldsheet:
\bea
\ddot{\textbf{x}} &=& \frac{1}{E}\left(\frac{\textbf{x}'}{Eh}\right)'
\label{xdamp}\\
\ddot{\eta} &=& \frac{1}{E} \left ( \frac{\eta'}{Eh}\right)' 
+ \frac{h_{,\eta}}{E^2h^2}
\left(\frac{3K^2}{\epsilon^{4/3}h}\textbf{x}'^2 + \eta'^2\right)
+ {\dot\eta}^2\left( \frac{K_{,\eta}}{K} - \frac{h_{,\eta}}{2h} \right)
\nonumber \\&&
-\frac{K_{,\eta}}{K} \frac{\eta^{\prime2}}{E^2h}
+ \frac{3K^2}{h} \left [
\left(Bh \right )_{,\eta} \dot{\phi}^2
- B_{,\eta} \frac {\phi'^2} {E^2} \right]
\label{etadamp}\\
\ddot{\phi} &=& \frac{1}{E}\left(\frac{\phi'}{Eh}\right)' 
+ \left(\frac{h_{,\eta}}{h}+\frac{B_{,\eta}}{B}\right)
\left(\frac{\phi'\eta'}{E^2h}-\dot{\phi}\dot{\eta}\right) \,.
\label{phidamp}
\eea
The quantity $E$ that appears in these equations is given by:
\begin{equation} \label{Efull}
E = \sqrt{\frac{-X'^2}{h\dot{X}^2}} = \sqrt{\frac{\textbf{x}'^2
+h\epsilon^{\frac{4}{3}}\left(\frac{\eta'^2}{6K^2}+B\phi'^2\right)}
{h\left(1-\dot{\textbf{x}}^2
-h\epsilon^{\frac{4}{3}}\left(\frac{\dot{\eta}^2}{6K^2}+B\dot{\phi}^2\right)
\right) } } \;.
\end{equation}
From the equation of motion for $t$, (\ref{t-eom}), we have $\dot{E} = 0$, 
i.e.\ $E$ is a conserved quantity. It is related to the conserved energy 
of the system, $\mathcal{E}$, by:
\be 
\mathcal{E} = \mu\int E(\sigma) d\sigma \,.
\label{relateEE}
\ee

Given that $E$ is conserved, it is easy to draw some qualitative 
conclusions from (\ref{Efull}). Essentially, the numerator in (\ref{Efull})
is related to the length of the string (via integration) and the 
denominator (ignoring $h$) how relativistic the string is at a particular
time. As the string falls down the throat, $h$ increases sharply, thus
presuming the string is not highly relativistic all along its length,
the $X^{\prime2}$ term must increase to compensate. Thus a string
falling down the throat will grow in the noncompact directions, as
well as stretching out in the internal directions, and vice versa.
However, there is clearly no friction term in these equations of motion
to cause the string to fall and be confined at the tip of the 
throat; rather, it is free to `bounce' back up again.
We will see this in both a simple example, as well as a full integration.

In flat spacetime, it is easy to obtain analytic solutions to 
the equations of motion \cite{KT2}. In warped spacetime, 
however, the equations of motion (\ref{xdamp})-(\ref{phidamp}), 
which are partial differential equations in $t$ and $\sigma$, are 
coupled and non-linear, and it is very difficult to find an 
analytic solution. One exception to this is the rather
special case in which the string is precisely at $\eta=0$: the
tip of the throat. Here, motion along a single great circle is now
exactly of the form considered in \cite{CCGGZ,CCGGZ2}, where
the extra dimension is circular. Strings at the tip of the conifold 
have been considered in \cite{JBP}, although the trajectory considered
there is rather special as it occurs at a polar singularity and actually
corresponds to a circular loop boosted along its length to the speed
of light. However, in spite of being somewhat non-generic, angular
motion of the string at the bottom of the conifold can give
interesting consequences for the noncompact trajectories, \cite{TIP}.

In order to model the dynamics we first 
look at a simple loop trajectory that reduces the 
partial differential equations to ordinary differential equations. 
We then go on to deduce general properties of string motion 
from the conserved energy, and to plot numerical 
solutions to the full equations of motion.

\subsection{A simple Ansatz} \label{ansatzsec}

A simple set-up in which the loop is pointlike in the internal 
dimensions and circular in the external dimensions reduces the 
equations of motion to a set of ordinary differential 
equations in $t$. The string has the freedom to move around 
in the throat, and for its radius in the external dimensions 
to expand and contract. We parametrize it as follows:
\be 
\label{circleansatz}
x^{M} = \left(t,\rho(t)\cos\sigma,\rho(t)\sin\sigma,0,\eta(t),\phi(t)\right)\,,
\ee
where $\rho(t)$ is the radius of the loop in the external dimensions, 
and the internal coordinates, $\eta(t)$ and $\phi(t)$, depend only on time.

The time-independent quantity $E$, given by equation (\ref{Efull}), is 
now also independent of the spacelike worldsheet coordinate, $\sigma$, 
so it is an absolute constant:
\begin{equation} \label{E}
E = \sqrt{ \frac{\rho^2}
{h\left(1-\dot{\rho}^2-h\epsilon^{\frac{4}{3}}
\left(\frac{\dot{\eta}^2}{6K^2} + B\dot{\phi}^2 \right)\right)} } = E_0
\end{equation}
The equations of motion for $\rho$ and $\eta$ become:
\begin{equation}\label{ODE1}
\ddot{\rho} = -\frac{\rho}{E_0^2h}
\end{equation}
\begin{equation}\label{ODE2}
\ddot{\eta} = \frac{3K^2h_{,\eta}\rho^2}{\epsilon^{\frac{4}{3}}E_0^2h^3} 
+ \dot{\eta}^2\left(-\frac{1}{2}\frac{h_{,\eta}}{h}
+\frac{K_{,\eta}}{K}\right) 
+ 3K^2\dot{\phi}^2\left(\frac{h_{,\eta}}{h}B+B_{,\eta}\right) \,.
\end{equation}
There is now a conserved angular momentum in the throat:
\be
\label{L} J = \dot{\phi}hB = const.
\ee
This is a Hamiltonian system with the conserved quantity $E_0$ being 
the Hamiltonian. Rearranging (\ref{E}), using (\ref{L}), expresses 
the system as motion in an effective potential:
\begin{equation} \label{effectivepot}
\dot{\rho}^2 + \dot{R}^2 = 1 - \frac{\rho^2}{E_0^2h} 
- \frac{\epsilon^{\frac{4}{3}}J^2}{hB}
=1 - V_{{\rm eff}}(\rho,R) \,,
\end{equation}
where
\be
\label{confR}
dR =  \sqrt{\frac{h}{6}}\frac{\epsilon^{\frac{2}{3}}d\eta}{K}
\ee
is an alternative conformal radial coordinate. 
In this system, since $E$ is conserved, we have free motion with 
no energy being lost\footnote{In reality, of course, the string 
will be losing some energy, for example through gravitational 
radiation, which we model in section~\ref{gravradn}.}. 

We can now qualitatively see the effect of motion in the throat
from the form of this potential. First note that $E^2_0h$ sets the
scale of oscillations in $\rho$, thus as the string moves down the
throat, the overall size of the string in the noncompact directions
will increase as already noted.  Therefore a string moving down
from the UV will dump energy into our visible loop. However, it is clear
from the form of the equations of motion that that energy will not 
remain in our noncompact dimensions, but that the string will continue to 
move back up the throat leeching energy back, and continuing to oscillate. 
The timescales of these relative oscillations depend on the initial 
scale of $\rho$, and the compactification parameters. 

To get an estimate of motion (before
presenting some exact solutions), consider a very simple set-up.
Suppose that the loop has only radial motion in the throat, and is
very close to the tip (i.e. the energetically favoured potential minimum), 
then expanding the various functions in the metric near the tip, we obtain
\be \label{effpot}
V_{\rm eff} (\rho,R) = \frac{\epsilon^{8/3}}{2E_0^2 (g_sM\alpha')^2}
\frac{\rho^2}{I_0}
\left ( 1 + \frac{\epsilon^{4/3}}{3(g_sM\alpha')^2}
\frac{R^2}{I_0^2} \right )
\ee
where $I_0 = I(0) = 0.5699$ is the value of the integral (\ref{Iofeta}) 
at the tip of the throat. We can now see the hierarchy of scales: provided
$R \ll g_s M \alpha' \epsilon^{-2/3}$, $\rho$ will have harmonic motion
at a frequency $\omega_\rho \sim \epsilon^{4/3}/(g_s M \alpha' E_0)$. 
Since the scale of $\rho$ is set by the string length, $L$, which is 
of order the Hubble scale for current day cosmic strings, this frequency 
is of order $L^{-1}$, as might be expected for a cosmic string. 
However, superimposed on this general behaviour are oscillations 
in $R$, which have a frequency set by the compactification parameters:
$\omega_R \sim \epsilon^{2/3} / (g_sM\alpha') \gg \omega_\rho$.

This is of course a ``broad brush" behaviour, once $\rho$ becomes very
small, the timescale of motion in $R$ lengthens, and $R$ can 
potentially move further up the throat. In fact, there is no reason in
principle for the motion not to explore all regions of the 
effective potential. As this is a Hamiltonian system, it does not have 
attractors (regions to which trajectories converge over time), and 
therefore there is no mechanism by which a string could be 
completely confined at the tip of the throat. Indeed, for small $J$,
the potential is strongly `creased' at $\rho=0$, with a sharp valley
along the $R$-direction. This makes the motion very sensitive to initial 
conditions should the loop happen to hit this direction in phase space.
We see this in some of numerical integrations.

The picture with angular momentum is broadly similar, as unless $J$ is
improbably large, the term in  (\ref{effectivepot}) is generally small.
The main qualitative difference with angular momentum is that it
gives a hard bound on how far up the throat the string can move, since
\be \label{etamax}
\frac{\epsilon^{\frac{4}{3}}J^2}{h(\eta)B(\eta)} \leq 1 \,.
\ee

It is straightforward to solve these equations of motion numerically,
although for practical reasons, we cannot access the most physically
realistic huge hierarchies between the scale of the throat and the 
current cosmological cosmic string network scale. However, these
solutions can test the general understanding built up by analysing
the system as above. Figure \ref{fig:withJ} shows a
demonstration of the qualitative motion described above. Here, 
the rather improbable set of data has been chosen: $\epsilon =
0.1 \alpha'^{\,3/4}$, 
$g_sM = 100$, $\rho(0) = 10^4 \sqrt{\alpha'}$,
$\eta(0) = 0.25$ (implying $R(0) = 57.86 \sqrt{\alpha'}$). 
For vanishing angular momentum these
give the values $E_0 = 4.4$, $\omega_\rho = 1.1 \times 10^{-4}$, 
$\omega_R = 2.2 \times 10^{-3}$. While these values are clearly
small, they are obviously highly unrealistic, nonetheless, the
behaviour of the loop does broadly follow the description given 
above. In the external dimension the loop oscillates with approximately
the above frequency $\omega_\rho$, while in the throat the 
loop oscillates faster, and when $\rho$ becomes small, 
it `coasts', since the potential (\ref{effpot}) is effectively flat. 
Note how this behaviour leads to the loop moving occasionally 
further up the throat, where the above approximation
to the potential will break down -- even so, the broad brush behaviour
continues, although the internal oscillations seem to prefer being 
stronger than the initial conditions would suggest. 

One can see on figure \ref{fig:withJ} the exchange of energy between 
the two sectors, since when internal oscillations of the loop are larger, 
external ones are smaller, and vice versa. Figure \ref{fig:vel} 
shows that a similar exchange occurs in the velocities. Clearly, 
the internal velocity, $v_{\eta}$, becomes small when external 
velocity, $v_{\rho}$, is large (which corresponds to the radius 
passing through zero). Looking more closely, one can see that, 
in agreement with \cite{tasos2008} for open strings, the 3D velocity
is modulated by the high frequency internal oscillations.
\FIGURE{
\includegraphics[scale=0.38]{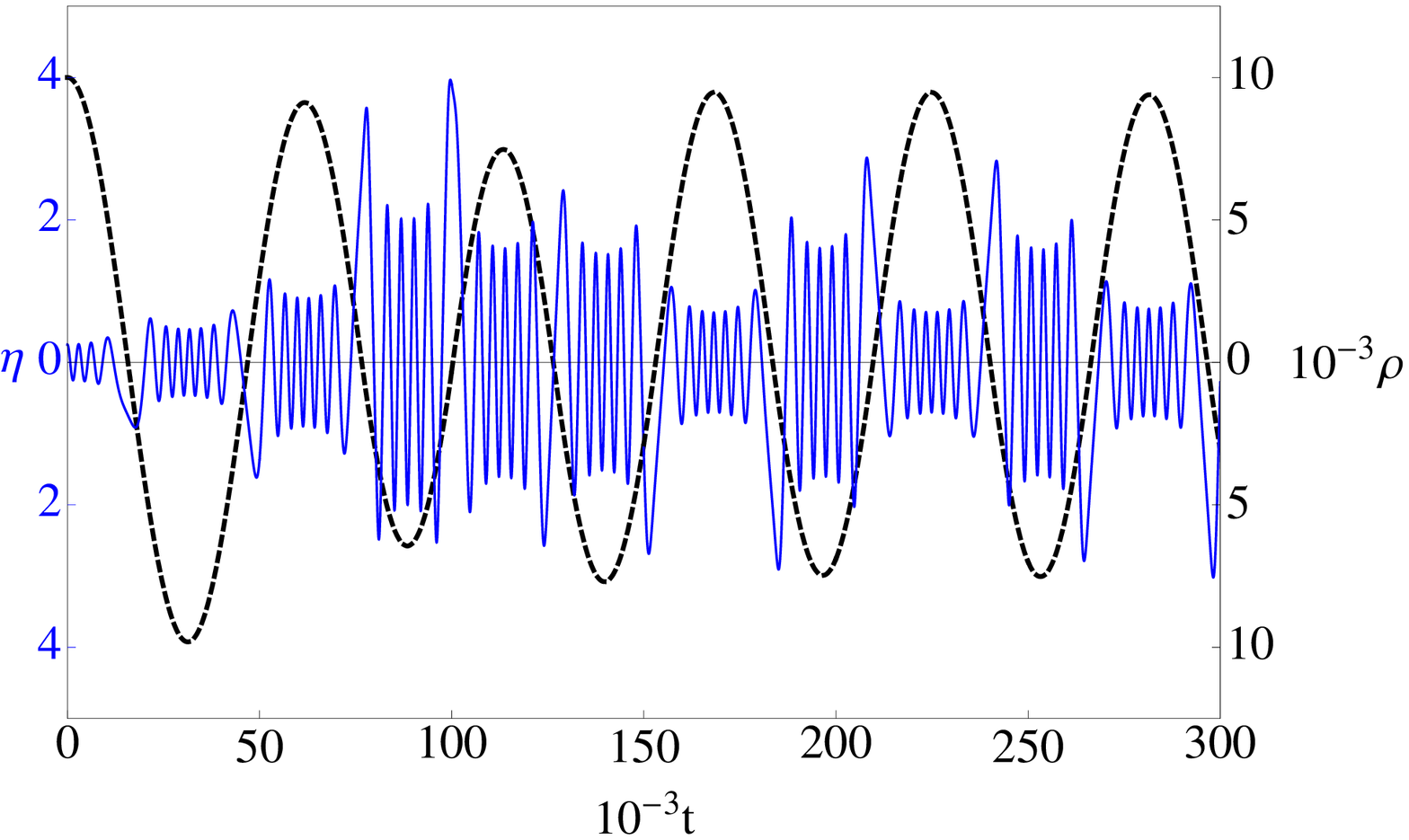}\;\nobreak
\includegraphics[scale=0.38]{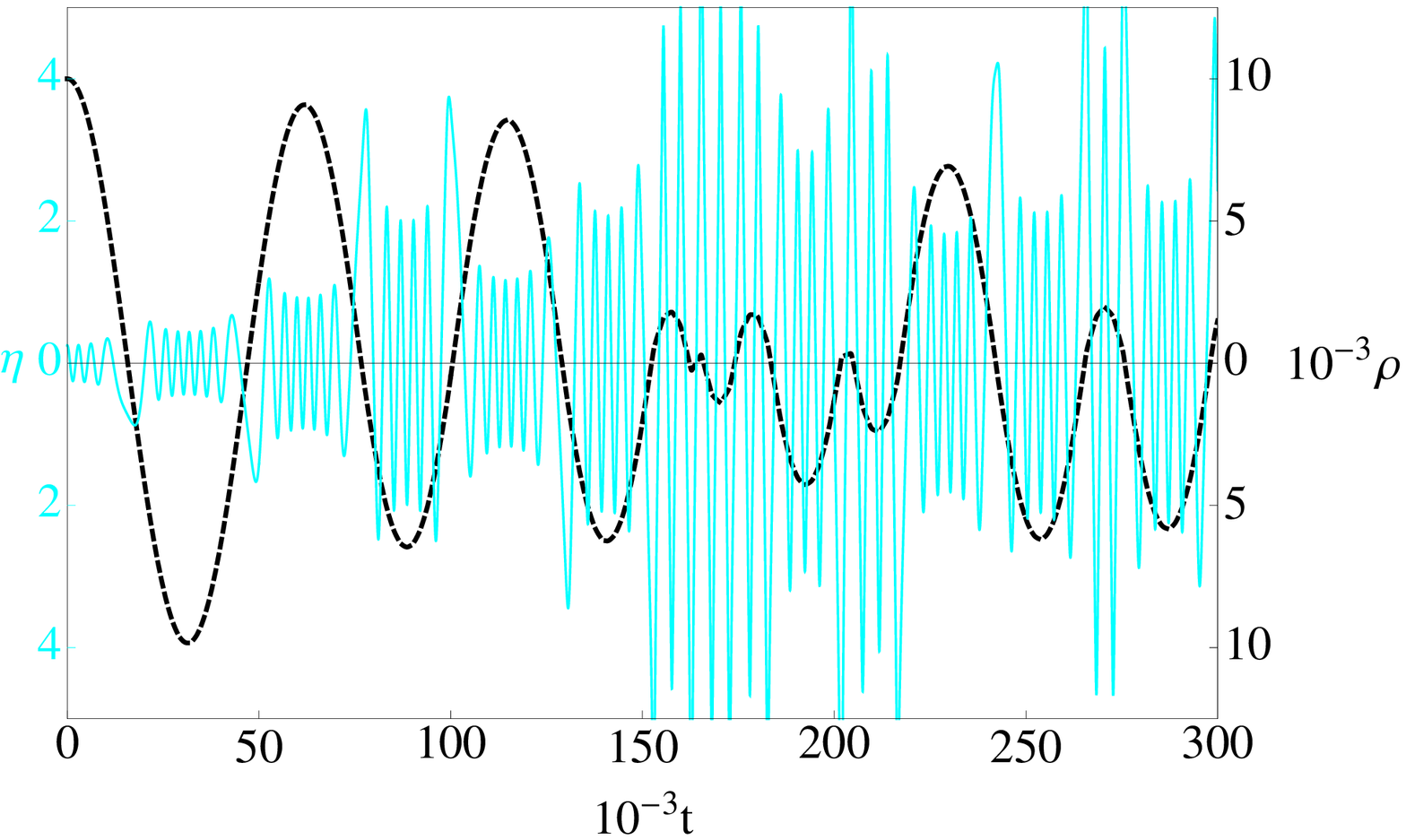}\\
\includegraphics[scale=0.38]{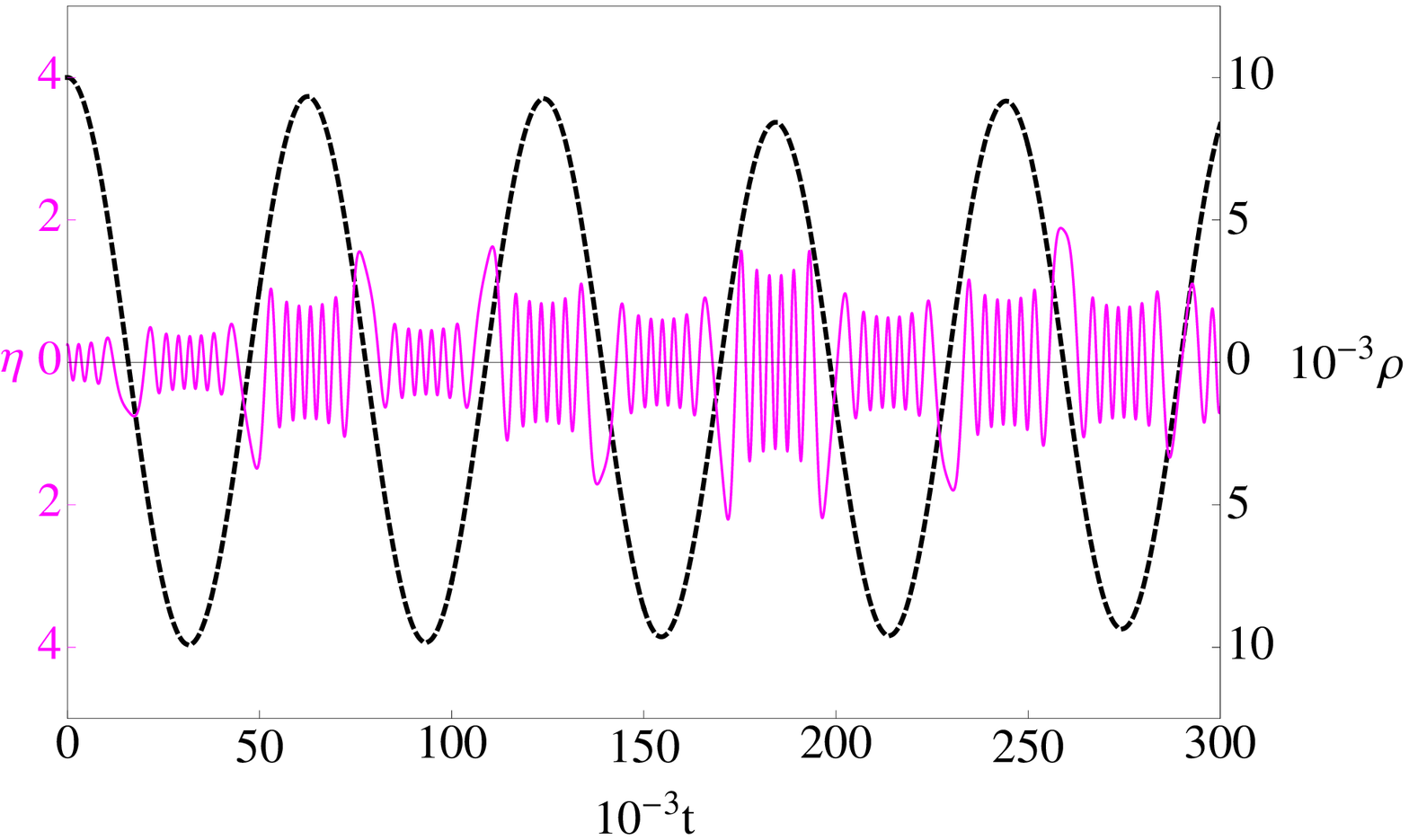}\;\nobreak
\includegraphics[scale=0.38]{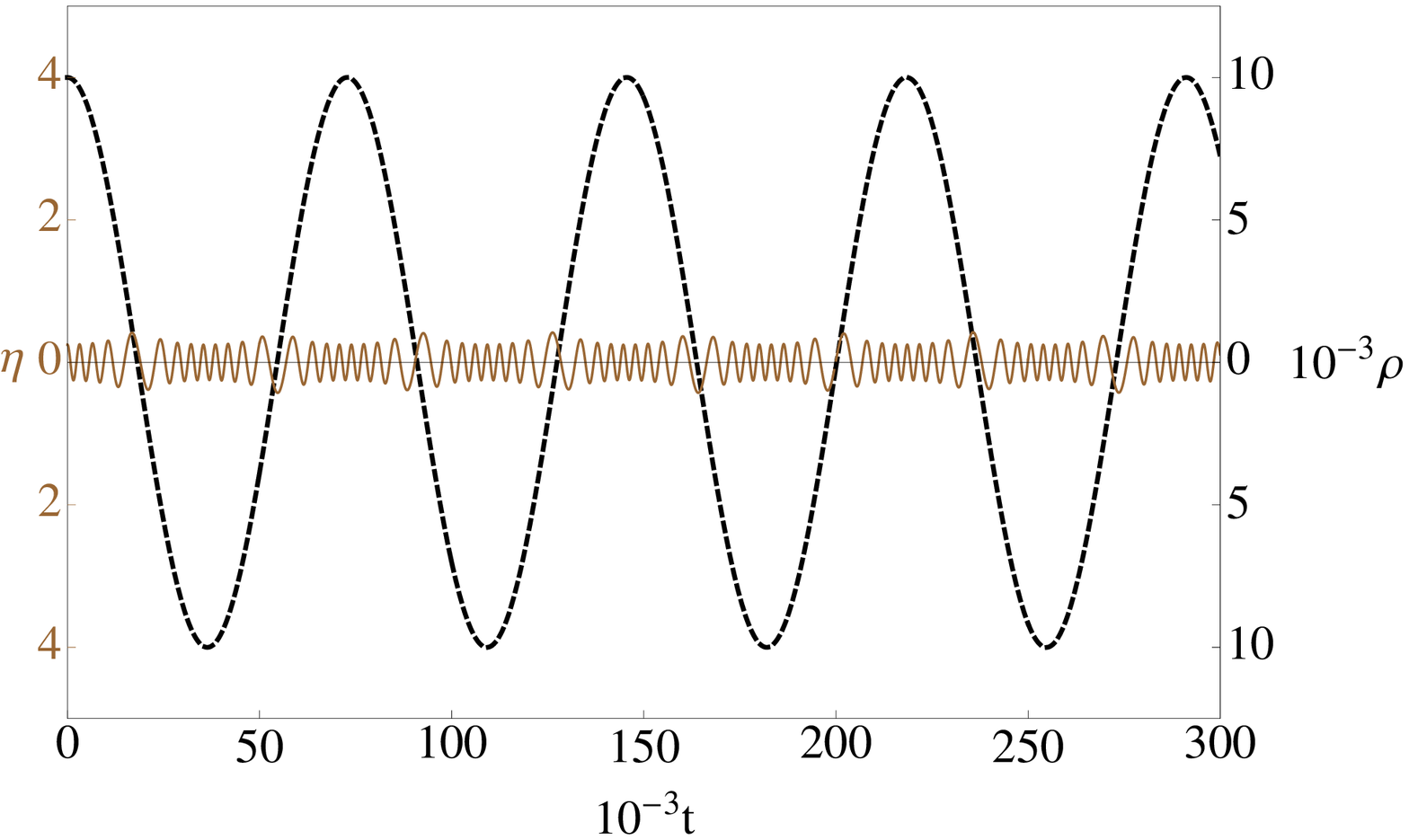}
\caption{Numerical solutions for the circular loop Ansatz (\ref{circleansatz})
showing the internal location $\eta(t)$ (in colours) and the external
radius $\rho(t)$ (in dashed black) of the loop, demonstrating
the interplay between external and internal motion and
the effect of angular momentum in the internal dimensions.
The plots show a loop of size $\rho_0 = 10^4\sqrt{\alpha'}$ initially
starting at rest at $\eta_0 = 0.25$, with $J = 0, 5, 10,$ and
$50$\% of the relativistic maximum $\epsilon^{-2/3} \sqrt{h(\eta_0)
B(\eta_0)}$, with $\eta(t)$ shown in blue, cyan, magenta, and
brown respectively. In all plots, the external loop radius, $\rho$,
is shown in dashed black.
}\label{fig:withJ} }
\FIGURE{
\includegraphics[width=0.47\textwidth]{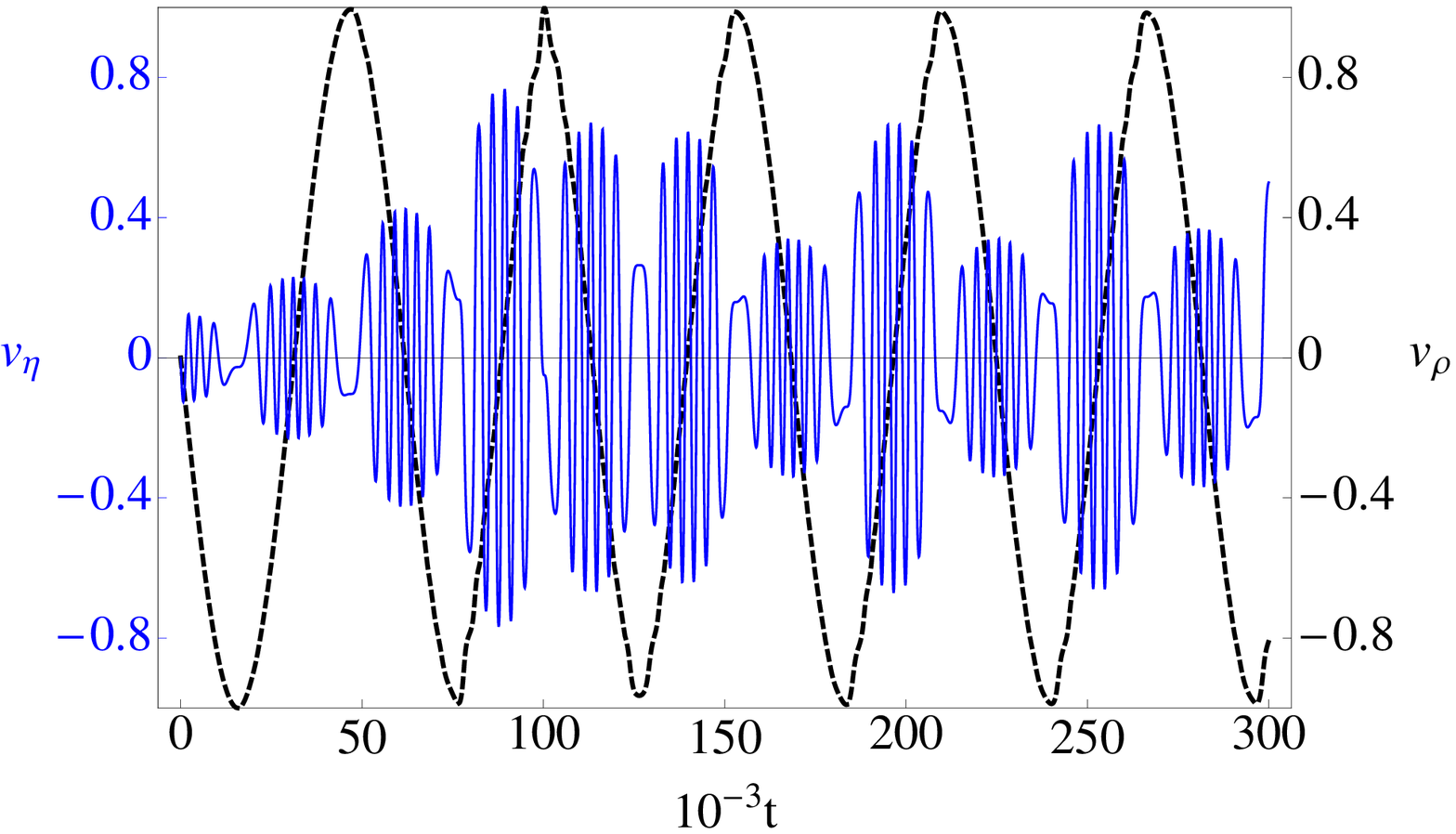}\ \ \ 
\includegraphics[width=0.47\textwidth]{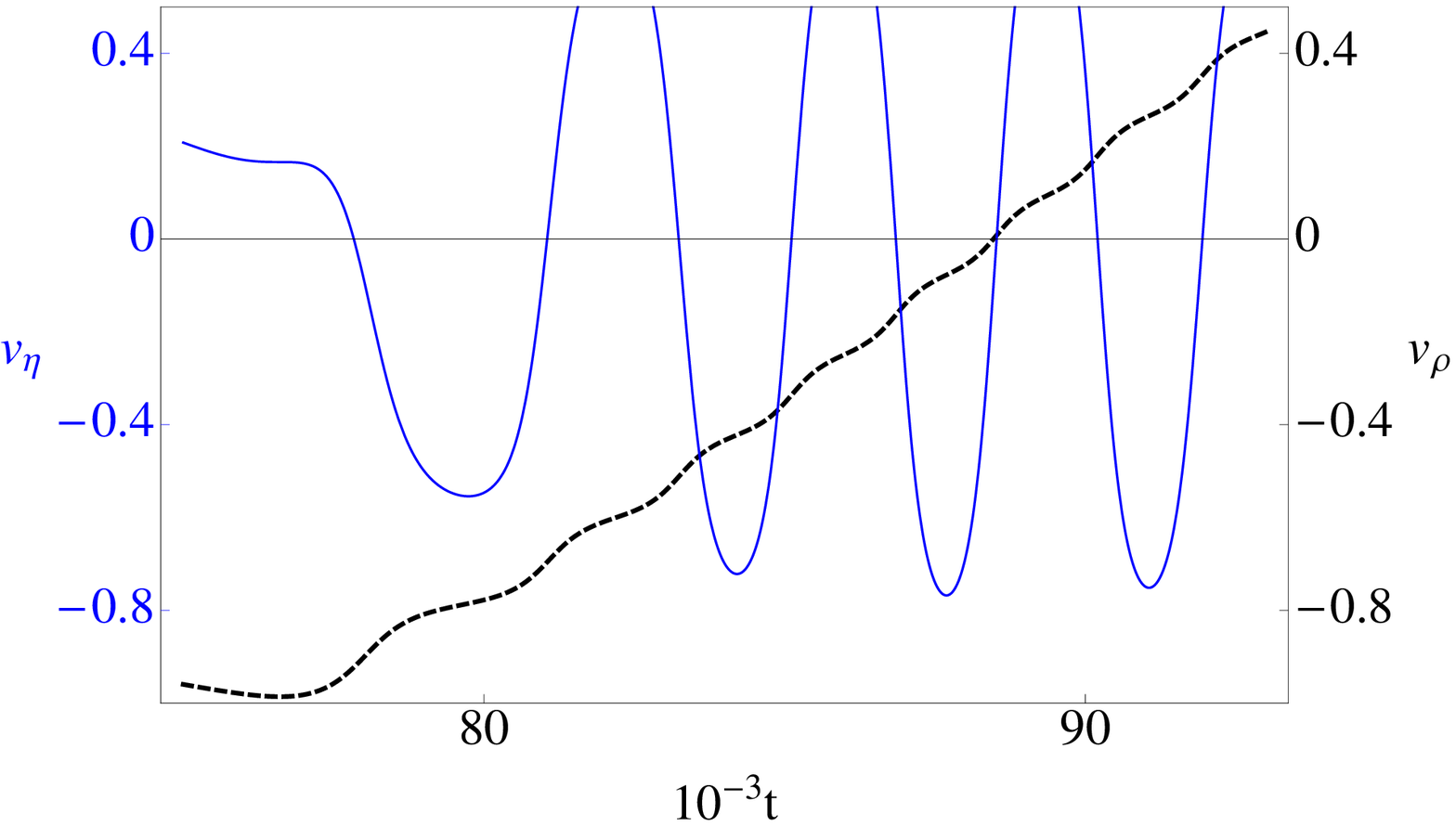}
\caption{Left: The physical velocities, $v_{\eta}$ (blue) 
and $v_{\rho}$ (dashed black), corresponding to the first plot 
of figure 1 ($J=0$). It is clear that energy is transferred back 
and forth between the two. Right: Detail from the left hand plot 
showing modulation of the 3D velocity, $v_{\rho}$ (dashed black), 
due to the higher frequency internal oscillations. }\label{fig:vel}}

The presence of an internal velocity is the important feature 
that in \cite{CCGGZ,CCGGZ2} was found to make a remarkable difference to the 
gravitational wave signal given off by string loops. 
The internal velocity causes the cusps on the string motion 
to be rounded off, and also to be less likely to form. 
This effect was measured by a ``near cusp event'' parameter 
$\Delta \simeq \sqrt{1-{\dot{\mathbf{x}}}^2} /2$, which we can 
compute by finding local maxima of the external velocity,
$|\dot{\mathbf{x}}|$, for these simple trajectories. For the loops
plotted in figure \ref{fig:withJ}, we find that $\langle\Delta\rangle=
0.071, 0.059, 0.099,$ and $0.252$ for $J=0, 5\%, 10\%,$ and $50\%$
of $J_{\rm max}$ respectively. The large value of $\langle\Delta\rangle$
for large $J$ is not surprising: in this case the loop has, 
by construction, a (conserved) relativistic internal motion. 
The values of $\langle\Delta\rangle$ for small, or zero, $J$ 
are more relevant, as these represent initial conditions where 
there is little relativistic motion internally.
Exploring further the parameter space for $J=0$ yields $\langle\Delta\rangle
= 0.012, 0.048,$ and $0.035$ for initial loop radii of $20$K$\sqrt{\alpha'}$, 
$30$K$\sqrt{\alpha'}$ and $40$K$\sqrt{\alpha'}$
respectively, indicating that this is not a parameter which
drops as the loop size is increased, but rather seems to respond more
to the interplay between the internal and external dimensions.

We conclude that in the presence of a warped throat, this 
particular loop trajectory has significant 
internal motion and cannot be approximated as effectively 4D.

\subsection{More general loops} \label{generalloops}

A general shape of loop will have much more complicated motion 
than the simple oscillations modelled in the previous section. 
There are potential terms in the equation of motion for $\eta$, 
equation (\ref{etadamp}), that include the derivative of the warp 
factor, $h$, for example the second term:
\be
\label{potterm} 
\frac{h_{,\eta}}{h}\frac{1}{E^2h}
\left(\frac{3K^2}{\epsilon^{\frac{4}{3}}h}\textbf{x}'^2 + 
\eta'^2\right)\,. 
\ee
Terms like this have the opposite sign to $\eta$ and therefore 
induce oscillations of the string up and down the throat. 
However, intuitively one expects the string's own tension to 
pull different parts of it in different directions, which, coupled with 
the forces from the warped throat, will result in complicated dynamics.

If, for a moment, we assume that the loop is very close to the tip of the
throat, we can do a similar expansion to that of (\ref{effpot}) for a more
general loop. 
If we first make the transformation $\sigma\rightarrow\tilde{\sigma}$
such that $d\tilde{\sigma} = E(\sigma)d\sigma$, then $\tilde{\sigma}$ 
runs from $0$ to $L$, the (10D) length of the loop, and the equation 
of motion for the external part of the loop (\ref{xdamp}) becomes:
\be 
\label{xKT} 
\ddot{\textbf{x}} = \frac{1}{h}\left(\textbf{x}'' 
- \frac{\textbf{x}'h_{,\eta}\eta'}{h}\right) 
\ee
where prime is now differentiation with respect to $\tilde{\sigma}$.

If we assume the external part of the loop is of cosmological size, 
it will contain most of the length of the string, i.e.\ $|\textbf{x}|
\sim {\cal O} (L)$. 
Then, assuming the loop can be approximated by low harmonics only, we also have
\be 
|\textbf{x}'|\sim {\cal O} (1) \;\;\;;\;\;\;\; 
|\textbf{x}''|\sim {\cal O} \left (\frac{1}{L}\right ) 
\;\;\;;\;\;\;\; \eta'\sim{\cal O} \left (\frac{\eta}{L}\right ) . 
\ee
Expanding $h$ and $h_{,\eta}$ in small $\eta$ gives the following:
\be  
h \simeq (g_sM\alpha')^2\epsilon^{-\frac{8}{3}}
\left(\frac{2}{3}\right)^{\frac{1}{3}}\left(0.72 - \frac{\eta^2}{3}\right) 
\hskip 5mm ; \hskip 1cm
\frac{h_{,\eta}}{h} \sim\eta \;.
\ee
This implies that 
$\frac{h_{,\eta}\eta'}{h}\sim\frac{\eta^2}{L}$, and thus that the second
term on the RHS of (\ref{xKT}) is ${\cal O}(\eta^2)$ suppressed relative
to the first. Thus if  $\eta\ll 1$, 
or $R \ll g_s M \alpha'\epsilon^{-2/3}$,
the overall factor of $\frac{1}{h}$ is approximately constant, and motion
will be close to a Kibble-Turok solution, for which
left and right-moving modes are independent, i.e.\ the effect of the internal 
motion will be small. Indeed, the form of this correction suggests that
corrections to the Kibble-Turok motion will be (roughly) of order
${\cal O}(\eta^2)$. Returning to our simple solutions of figure
\ref{fig:withJ}, computed with initial data $\eta_0 = 0.25$, this would
suggest a discrepancy of order $0.04$ from the exact 4D Nambu string,
which is the ballpark of the estimates for $\Delta$ from the
numerical solutions.

For a loop solution of the type (\ref{twod}) with significant and 
generic motion in the internal dimensions, however, a numerical 
integration of the PDE's (\ref{xdamp}-\ref{phidamp}) is necessary to find the 
detailed and explicit trajectory. To explore the effect of more complex 
internal motion, we started with a circular loop in the external space, 
as in the simple Ansatz in the previous section, but instead of
the ``pointlike'' internal motion of our previous Ansatz, we allowed for a 
loop structure in the internal dimensions as well. Note, that even though the 
initial loops (both in the external and internal dimensions) were set up 
to be circular, there is now no symmetry condition imposed on the 
solution at $t>0$. Figures \ref{pdetraj1}
and \ref{pdetraj2} show some sample snapshots of the loop's motion
once the evolution is underway.
In these plots, for numerical expedience the 
more unrealistic compactification parameters, $g_sM = 10$
and $\epsilon = 0.5\alpha'^{3/4}$ were used, and the radii of the internal 
and external extent are more similar.
\FIGURE{
\includegraphics[scale=0.4]{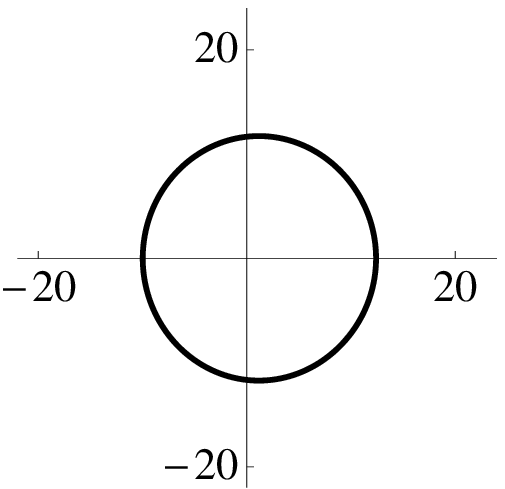} \includegraphics[scale=0.4]{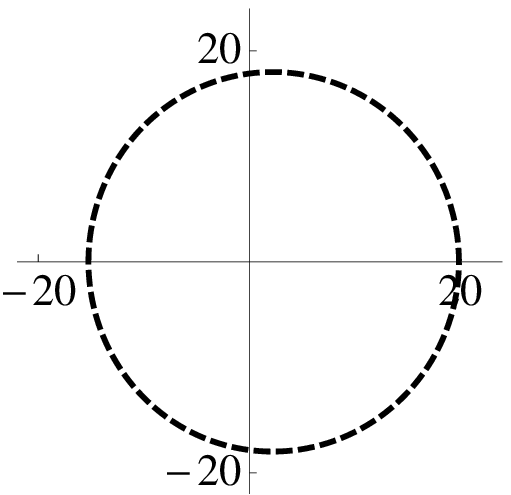} 
\includegraphics[scale=0.4]{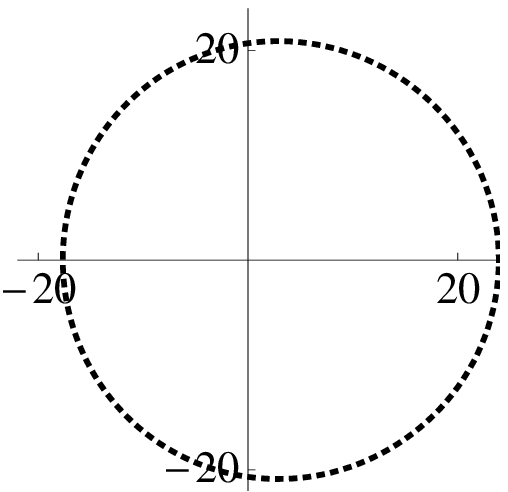} \includegraphics[scale=0.4]{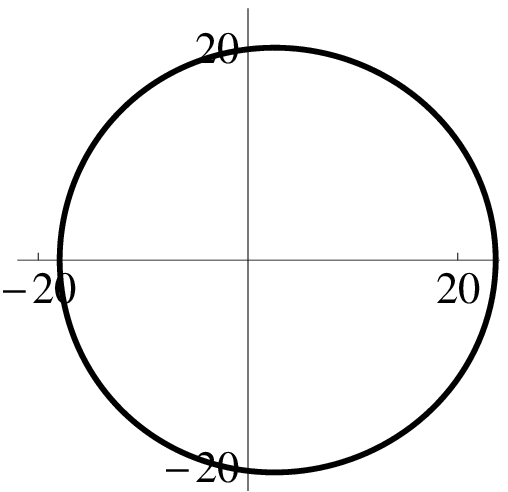} 
\includegraphics[scale=0.4]{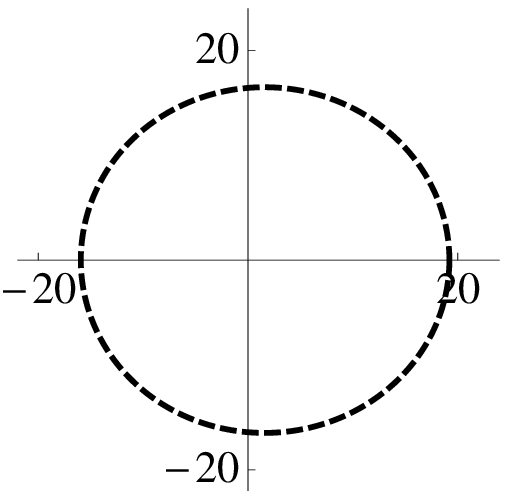} \includegraphics[scale=0.4]{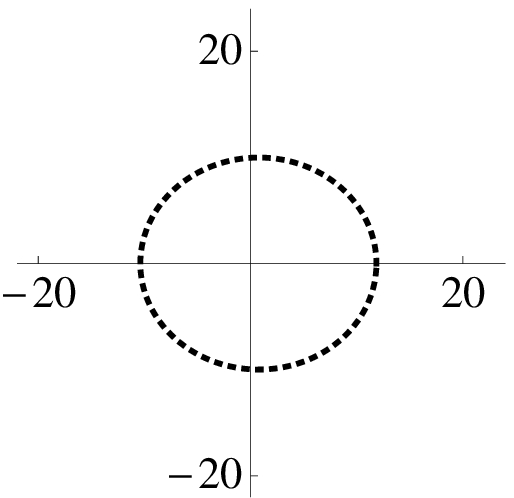}
\includegraphics[scale=0.4]{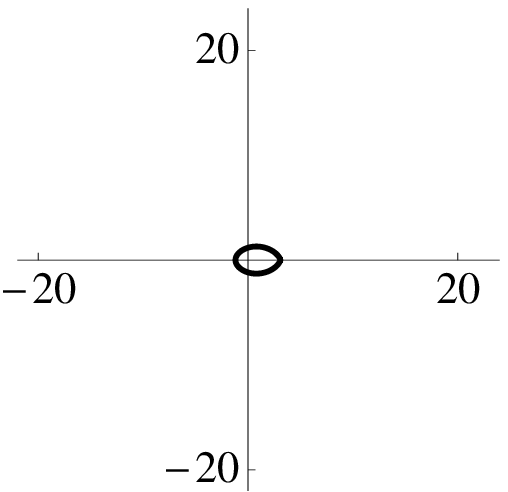}
\includegraphics[scale=0.5]{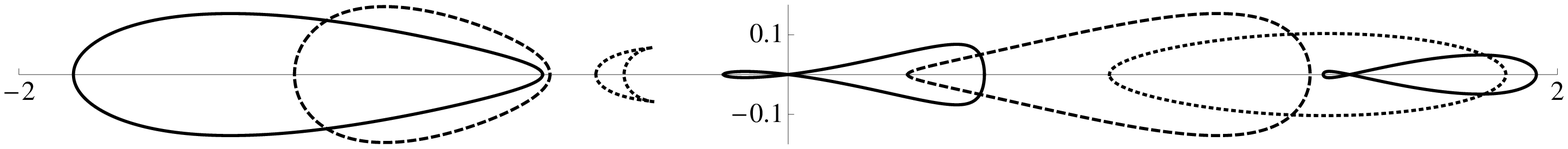}
\caption{Snapshots of a string loop evolving both externally (upper) and 
internally (lower), where the initial internal loop is displaced up the throat. 
In the lower plot, the axes are $\eta\cos\theta$ and $\eta\sin\theta$,
and the snapshots are selected some time after the start of the integration.
}\label{pdetraj1} }

The plots show how, broadly, the features that were demonstrated in
the simple Ansatz are also present here: the loop oscillates up 
and down the throat with periods when motion is concentrated closer to 
the tip translating into a larger, or more relativistic, loop in the 
external dimensions (and vice versa).
Figure \ref{pdetraj1} shows an example with the loop starting
up the throat, moving down and back up again, while also
changing shape under its own tension. 
Although the loop is clearly no longer pointlike in the internal 
dimensions, it is fairly localised, making the trajectory close 
to that of the simple ansatz. 
Indeed a similar behaviour is seen to occur, with the external part of the
loop remaining roughly circular and oscillating in and out, its maximum 
radius of oscillation depending on how close to the tip of the throat 
the loop is at the time (which we see by observing the trajectory 
over longer timescales).  A fairly coherent oscillation of the 
whole loop up and down the throat is also observed.  This consistency 
of the full system with the simple ansatz reinforces the conclusions 
made in section \ref{ansatzsec}. Notably, on figure \ref{pdetraj1}, 
the loop becomes larger as the internal loop contracts. 
Although in a realistic cosmological setting, there would be a large 
hierarchy between internal and external frequencies, we would 
still expect to see a similar exchange of energy (as in 
figure \ref{fig:withJ}, for example), averaged over internal oscillations.

Figure \ref{pdetraj2} focuses instead on a loop whose initial
internal configuration encircles the origin, and has a significant
variation in $\eta$. Here, the part of the loop at larger $\eta$ 
starts to fall down the throat, whereas the part at lower $\eta$ begins 
to move up.  As the loop then has varying motion internally, this
feeds into the external motion, causing the loop there to start curling,
forming (apparently) a double loop, then a very wrinkled loop. This
particular example shows a dramatic change of shape in the external
dimensions, which we expect will be muted as the radius of the loop
in the external dimensions is increased to more realistic
cosmological scales.
\FIGURE{
\includegraphics[width=0.16\textwidth]{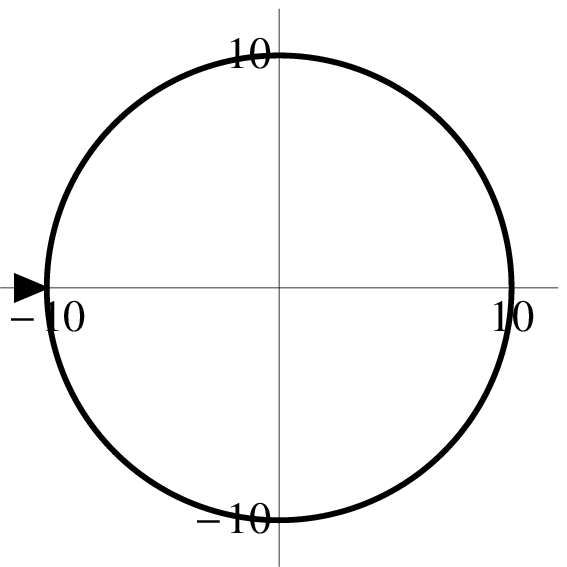}\nobreak
\includegraphics[width=0.16\textwidth]{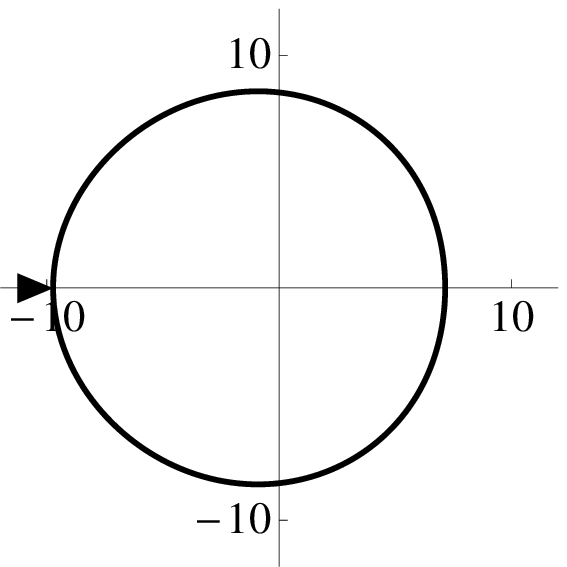}\nobreak
\includegraphics[width=0.16\textwidth]{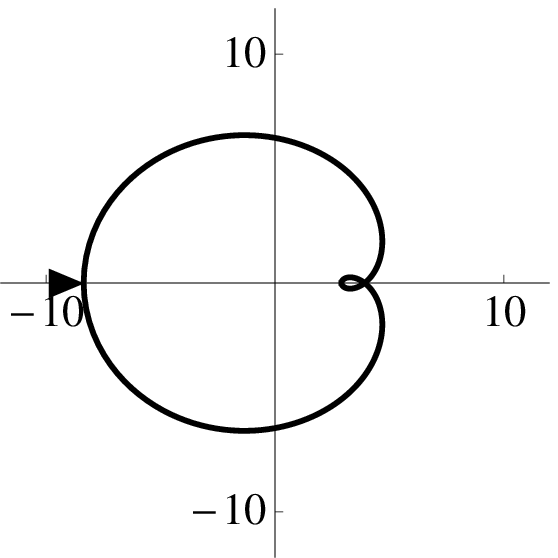}\nobreak
\includegraphics[width=0.16\textwidth]{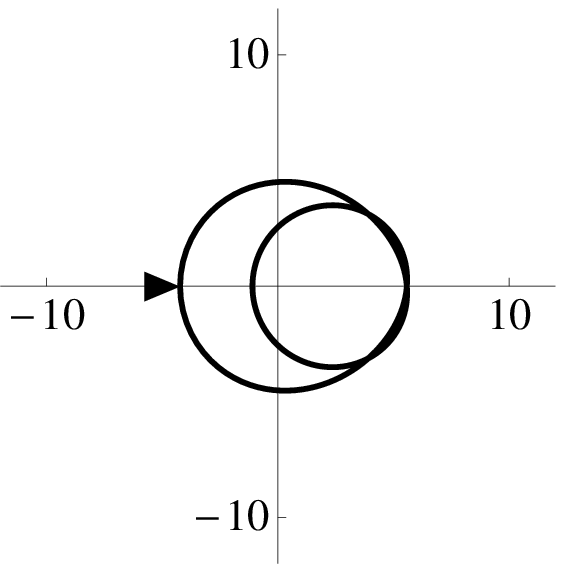}\nobreak
\includegraphics[width=0.16\textwidth]{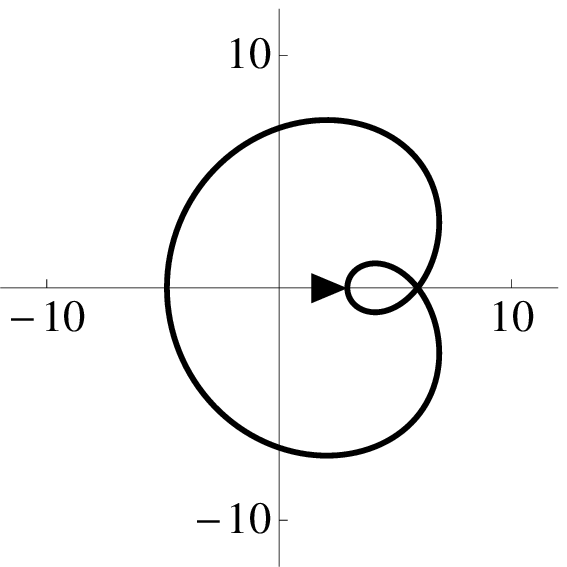}\nobreak
\includegraphics[width=0.16\textwidth]{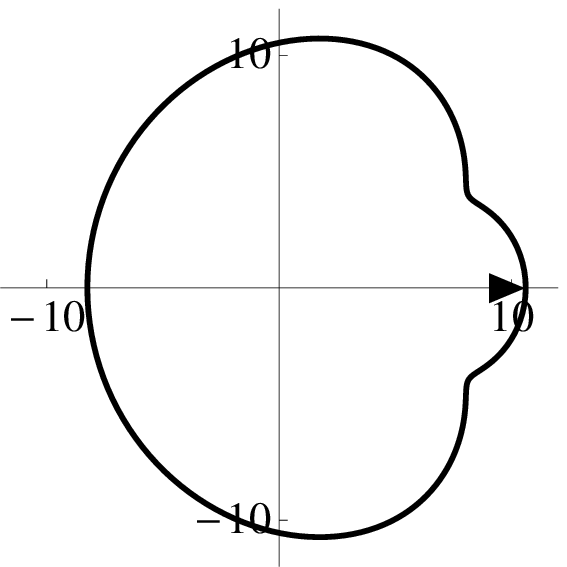}\\
~\includegraphics[width=0.16\textwidth]{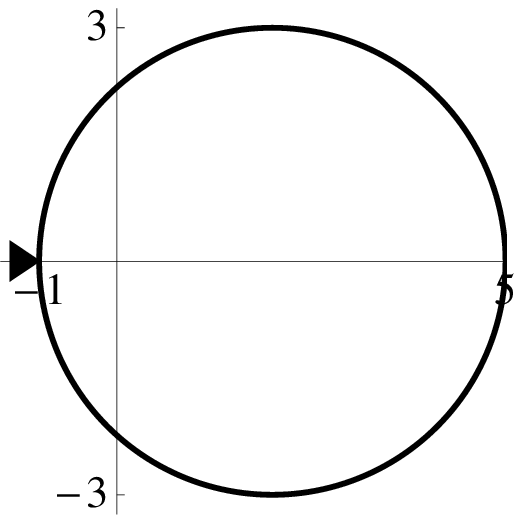}\nobreak
\includegraphics[width=0.16\textwidth]{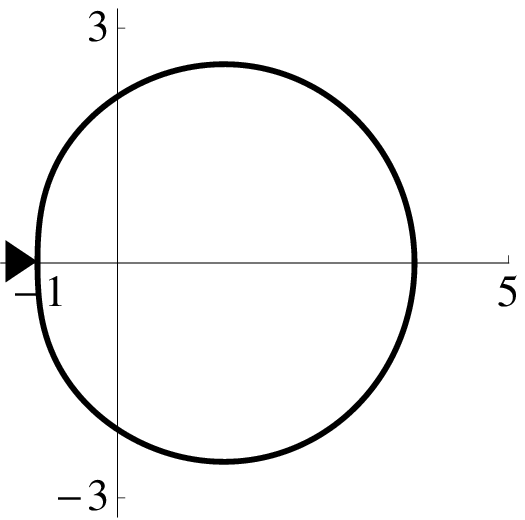}\nobreak
\includegraphics[width=0.16\textwidth]{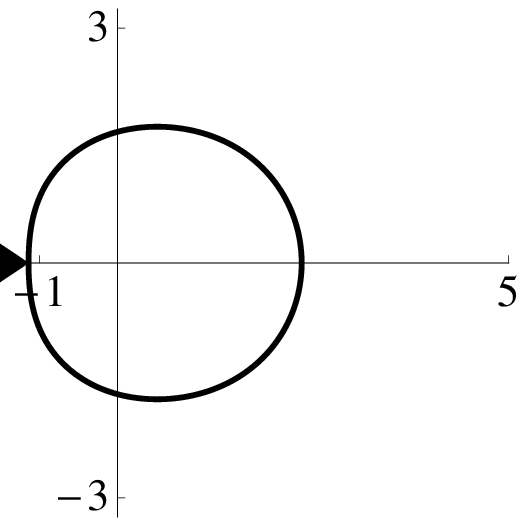}\nobreak
\includegraphics[width=0.16\textwidth]{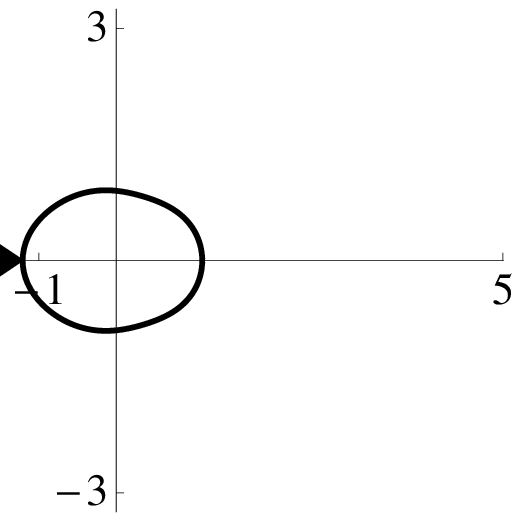}\nobreak
\includegraphics[width=0.16\textwidth]{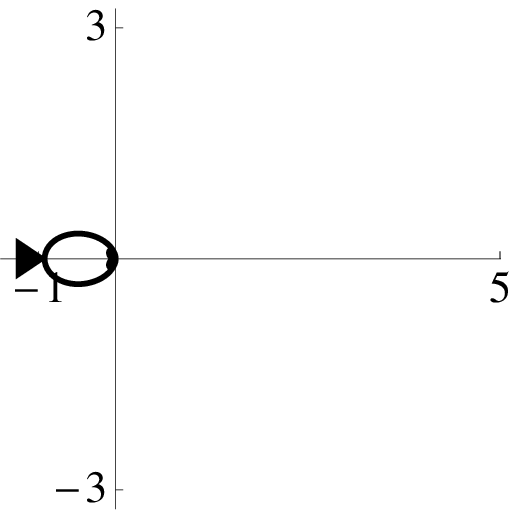}\nobreak
\includegraphics[width=0.16\textwidth]{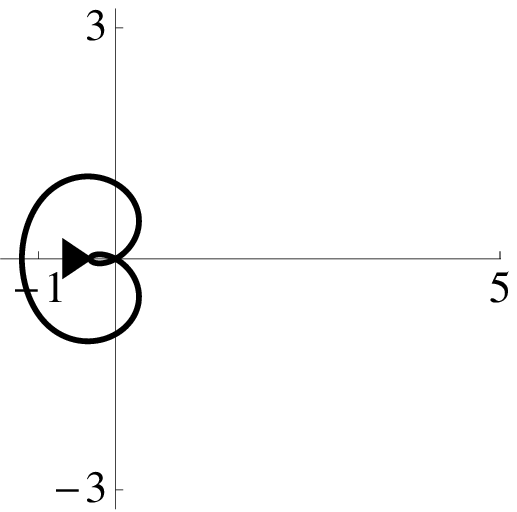}
\caption{From left to right, snapshots of a string loop at 
evenly spaced time intervals, the upper sequence showing the loop
in the external dimensions, and the lower sequence in the internal
dimensions. This solution demonstrates how the circular shape 
is distorted by internal motion, and develops apparent kinks
and crossings (although the loop does not self intersect as
it misses in the internal dimensions). The arrow denotes 
the position on the loop where $\sigma=0$ for comparison
of the location of the features.
}
\label{pdetraj2} }

It is interesting to note that even for a finite energy, $\mathcal{E}$, 
there is no hard bound on how far from the tip of the throat a 
string can move (i.e. how large $\eta$ can be). While we can bound
the variation in $\eta$ at a particular time ($\mathcal{E} \geq
\mu \epsilon^{2/3} \int d\sigma |\eta'|/\sqrt{6}K = 2\mu 
\epsilon^{2/3} \Delta R(t)$), we can nonetheless have the loop
high up the throat if we effectively `freeze' it in the external
dimensions (i.e.\ temporarily have the loop very small and nonrelativistic).
Such a situation however, would require a conspiracy of internal
and external motion that makes it unlikely, or a rare event.
More generically, we expect a loop to approximately remain within 
an order of magnitude or so of its inital data (in the absence of 
cosmological evolution) as it is unlikely for a loop with general internal
shape to move significantly up the throat. Even this however,
is enough to maintain sufficient internal motion, and does not
preclude significant local alterations, such as the quirks exhibited
in figure \ref{pdetraj2}.

\subsection{Effect of two-form charges} \label{bfield}

In many explorations of cosmic superstrings, the strings are
considered to interact with the IIB supergravity fields, in
particular the NS-NS 2-form (for the fundamental string) or
the RR 2-form (for the D-string), or indeed both (for $(p,q)$
strings). While it is unclear to what extent these supergravity
interactions remain at low energies, since the exact solution
we are using for the throat contains fluxes of both these
fields, we should be mindful of the effect of these 2-forms
on the motion of the strings. A fully general solution will be
rather involved, however, we comment here on general features
of including these terms in the motion of the string.
For a string charged under a 2-form, the Nambu action acquires
an additional term:
\be
S \propto -\int d\sigma dt \sqrt{-\gamma} 
+ \int d\sigma dt\epsilon^{AB}X^{a}_{,A}X^{b}_{,B}B_{ab} \,. 
\ee
The equation of motion is then:
\be 
\label{Beom}
\begin{aligned}
\frac{1}{\sqrt{-\gamma}} \left [ \frac{\partial}{\partial t}
\left(\frac{\dot{X}^{a}X'^{2}}{\sqrt{-\gamma}} \right) + 
\frac{\partial}{\partial \sigma}
\left(\frac{X'^{a}\dot{X}^{2}} {\sqrt{-\gamma}}\right) \right ] + 
& \frac{1}{\gamma} \Gamma^{a}_{bc}\left( X'^{2} \dot{X}^{b}\dot{X}^{c} 
+ {\dot X}^2 X'^{b}X'^{c} \right) \\
&= ({\dot X}^b X'^c - X'^b {\dot X}^c)H^{a}_{\ bc} \,.
\end{aligned}
\ee
Here we will consider explicitly $H_3 = dB_2$ from the 2-form (\ref{Bflux}),
as this will capture the essential qualities of including charge in the string.
(A D-string couples differently to the fluxes, but the impact will be
qualitatively the same.)

The key feature of including charge for the string is that, just like
the path of a charged particle is curved in a magnetic field, the charged
string will pick up transverse forces coming from the $H_3$ or $F_3$ fluxes.
Thus, we do not expect a simple single-angle motion, such as that of
the simple Ansatz considered in (\ref{twod}). In fact, if one sets 
$\theta_1=\theta_2 = \pi/2 = -\psi/2$, and $\phi_1 = \phi_2 
= \phi(t,\sigma)$, then we have a nonzero `force'
in the $\theta_i$ directions given by
\be
F^{\theta_1} = - F^{\theta_2} = 4g_sM \epsilon^{-4/3} \frac{k'(\eta)}
{\sqrt{h}\,B(\eta)} ({\dot\eta}\phi' - \eta'{\dot\phi})\; ,
\ee
where $k'(\eta)$ is the derivative of the function premultiplying 
$g^3\wedge g^4$ appearing in (\ref{Bflux}). 
This shows that the general motion will be quite involved. However, clearly
at least one internal coordinate must have $\sigma$ dependence 
as well as time dependence in order for this effect to show up, so
in fact the simple time dependent only Ansatz used in Section 
\ref{ansatzsec} is still a valid solution. 

Since there is no $H^{0}_{\ ab}$ component, the quantity $E$ 
given by the Nambu dynamics is still conserved, so the system 
still has a conserved energy. In order for the B-field to localise the string 
at the bottom of the throat, therefore, energy would have to be 
transferred into the other dimensions. 
This would require a damping term in the equation of motion 
for $\eta$, ie. a term with opposite sign to 
$\dot{\eta}$. The extra terms in the equation of motion 
for $\eta$ that are induced by the B-field are 
(from (\ref{Beom})):
\be
({\dot X}^b X'^c - X'^b {\dot X}^c) g^{\eta\eta} H_{\eta bc} \,.
\ee
Since $H_3$ is antisymmetric, both $b$ and $c$ must be 
angular directions, so there can be no 
term containing $\dot{\eta}$, and therefore no damping term.

One interesting piece of information that can explicitly be 
extracted from the B-field is the behaviour at 
the tip of the throat. In the case of purely Nambu dynamics, if 
a string is placed at rest at the tip of the 
throat, the Nambu equation of motion for $\eta$ (the radial coordinate) 
will vanish and it will simply stay 
there. Although one would not expect a general trajectory to 
reach this configuration, it is a valid 
solution. In the presence of the B-field, however, this 
is no longer the case. This stems mainly from the 
fact that part of the conifold geometry does not shrink to 
zero size at the tip of the throat, so it is still 
possible to have angular dependence and motion. In fact the 
geometry at the tip of the throat is a round 
3-sphere, which can be parametrized by setting $\theta_1 = \phi_1 = 0$, 
so the coordinates on the $S^3$ are $\theta_2$, $\phi_2$ and $\psi$. 

By calculating the relevant terms in $H_3$, we find that the 
following non-zero term appears in the 
equation of motion for $\eta$ at $\eta=0$:
$$
\left(\phi'_2\dot{\theta}_2 - \dot{\phi}_2\theta'_2\right)
h^{\frac{1}{2}}(0)\epsilon^{-\frac{4}{3}}(g_sM \alpha')K^2(0)\sin\theta_2
\sim \left(\phi'_2\dot{\theta}_2 - \dot{\phi}_2\theta'_2\right)\sin\theta_2 \,.
$$
Thus, if a string has both non-trivial extent and velocity on the 3-sphere, 
the presence of the B-field provides a force in the $\eta$ direction and 
pushes the string away from the origin, which further 
supports our argument that it will, in general, not be localised there.

Finally, since the interaction with fluxes clearly increases the amount
of internal motion, the effect on the external loop motion will
be to make it deviate even more from the exact Kibble-Turok form.

\section{Cosmological loops} \label{moreeffects}

We have argued that the motion of loops in the presence of 
warped extra dimensions will include all dimensions, and will 
not be localised at the tip of the throat. This was based 
on the assumption of classical dynamics in a static spacetime. 
Potentially significant modifications to this behaviour come from a 
variety of factors. In the Nambu equations of motion we saw 
potential terms that caused the string to 
oscillate about the tip of the warped throat, but 
argued that for it to be confined at the bottom, it would 
have to lose energy in the internal dimensions, which would 
require some kind of friction. Cosmological 
expansion causes damping of long strings in a 
network \cite{VOS,kibble85}, and therefore may provide 
this internal friction. Energy loss via emission of 
gravitational radiation must also be considered, since if 
a string were to lose more energy in the internal dimensions 
than the external ones this could result in 
an effectively 4D motion. 
We now consider both of these effects 
and find that our conclusions remain unchanged.

\subsection{Cosmological expansion} \label{expansion}

In an expanding universe, the velocities of long strings are damped, 
allowing the network to reach a scaling solution \cite{VOS,kibble85}. 
Thus cosmological expansion might prove to be a source of damping in the
internal dimensions, which might result in strings being confined
to the bottom of the warped throat. The important quantity to consider 
is the relative magnitude of internal and external damping, since a stronger 
damping in the internal dimensions would result in internal motion 
being effectively brought to a standstill, whilst the external 
part of the loop would continue to evolve, giving an effectively 
4D motion. In \cite{tasos2008}, this was explored for long
strings, and it was found that internal damping is in fact very weak, 
and does not localise the strings at the tip of the throat.

The effect of expansion on closed loops is more interesting. 
While outside the horizon scale, they behave similarly to
long strings: expansion effectively has a ``stretching'' effect, and
reduces the velocity of the string. For the closed loop, stretching
increases its total energy, and does slow down the motion of the 
loop for a while, but eventually the tension of the loop causes it
to contract and fall inside the horizon. Once the loop is well inside 
the horizon, expansion ceases to affect its motion. The transition 
between these two stages, when the loop is comparable to the horizon 
size, can have interesting dynamics.

For an FRW universe, the metric now takes the form:
\be
\label{metric}
ds^2 = h^{-\frac{1}{2}}(dt^2 - a^2d\textbf{x}^2) 
- h^{\frac{1}{2}} {\tilde g}_{mn} dy^m dy^n \,,
\ee   
so that the physical distance in the external dimensions is 
now $a\textbf{x}$. The scale factor, $a$, can be taken to be 
proportional to $t^{\beta}$, where $\beta = \frac{1}{2},\frac{2}{3}$ 
respectively in the radiation and matter eras. The scalar 
$E\equiv \sqrt{\frac{-X'^2}{h\dot{X}^2}}$, which in flat spacetime 
was conserved, now depends explicitly on the scale factor $a=a(t)$: 
\be 
\label{Eoft}
E = \sqrt{\frac{a^2\textbf{x}'^2+h\epsilon^{\frac{4}{3}}
\left(\frac{\eta'^2}{6K^2}+B\phi'^2\right)}
{ h\left(1-a^2\dot{\textbf{x}}^2-h\epsilon^{\frac{4}{3}}
\left(\frac{\dot{\eta}^2}{6K^2}+B\dot{\phi}^2\right)\right)}}\,.
\ee
The equations of motion become:
\bea
\dot{E} &=& -E\frac{\dot{a}}{a}\left(a^2\dot{\textbf{x}}^2
-\frac{a^2\textbf{x}'^2}{E^2h}\right)
\label{Edamp2}\\
\ddot{\textbf{x}} &=& \frac{1}{E}\left(\frac{\textbf{x}'}{Eh}\right)'
- \dot{\textbf{x}}\frac{\dot{a}}{a}\left(2-a^2\dot{\textbf{x}}^2
+\frac{a^2\textbf{x}'^2}{E^2h} \right)
\label{xdamp2}\\
\ddot{\eta} &=& \frac{1}{E}\left(\frac{\eta'}{Eh}\right)' 
+ \frac{h_{,\eta}}{h}\frac{1}{E^2h}\left(\frac{3K^2a^2}
{\epsilon^{\frac{4}{3}}h}\textbf{x}'^2 + \eta'^2\right)
+ \dot{\eta}^2\left(\frac{K_{,\eta}}{K}-\frac{h_{,\eta}}{2h}\right)
- \frac{K_{,\eta}}{K}\frac{\eta'^2}{E^2h}
\nonumber\\
&&+ 3K^2\dot{\phi}^2\left(B_{,\eta}+\frac{h_{,\eta}}{h}B\right)
- 3K^2B_{,\eta}\frac{\phi'^2}{E^2h} 
+ \dot{\eta}\frac{\dot{a}}{a}\left(a^2\dot{\textbf{x}}^2
-\frac{a^2\textbf{x}'^2}{E^2h}\right)
\label{etadamp2}\\
\ddot{\phi} &=& \frac{1}{E}\left(\frac{\phi'}{Eh}\right)' 
+ \left(\frac{h_{,\eta}}{h}+\frac{B_{,\eta}}{B}\right)
\left(\frac{\phi'\eta'}{E^2h}-\dot{\phi}\dot{\eta}\right)
+ \dot{\phi}\frac{\dot{a}}{a}\left(a^2\dot{\textbf{x}}^2
-\frac{a^2\textbf{x}'^2}{E^2h}\right) \,.
\label{phidamp2}
\eea
We see directly from (\ref{Edamp2}) that the quantity $E$  in (\ref{Eoft}), 
and therefore the total energy, $\mathcal{E} = \int d\sigma E$, is no 
longer conserved, as anticipated from its scale factor dependence. 
A straightforward consequence of (\ref{Eoft}) is that $-1 \leq 
\left(a^2\dot{\textbf{x}}^2-\frac{a^2\textbf{x}'^2}{E^2h}\right) \leq 1$,
and hence
\be
|\dot{E}| \leq E \frac{\dot a}{a} \,.
\ee
From this it follows that in order for $\dot{E}$ to be significant, $E$ 
must be a sizeable fraction of the horizon scale, i.e.\ 
$E\sim \left(\dot a/a\right)^{-1} = H^{-1}$. Physically, $\dot{E}\sim HE$ 
corresponds to a large, non-relativistic loop, where the energy 
is given by its rest-mass (length) and the dominant mechanism in 
(\ref{Edamp2})  is conformal stretching, increasing the total energy.  
On the other hand,  $\dot{E}\sim -HE$ corresponds to an ultra-relativistic 
loop, for which the relevant mechanism is velocity redshifting 
in the directions transverse to the string, decreasing the energy.

Useful insights into the behaviour of string loops in the spacetime 
(\ref{metric}) can be gained by focussing on the quantity $E$.  
Table \ref{expansiontab} details extreme cases of string motion 
(i.e.\ cases in which the dynamics is dominated by certain types of terms), 
and the corresponding behaviour of $E$. We expect, and indeed find 
numerically, that energy is generally transferred back and forth 
between length and velocity (columns 1 and 2) and the
extra dimensional motion (column 3).  This implies that $E$ 
can attain its limiting behaviour -- increasing proportionally to 
the scale factor -- only part of the time. In general, its growth will be 
slower, and, at certain intervals during the loop's evolution, $E$ 
can even decrease.  On the other hand, the scale factor, $a$, will 
always be increasing as $t^\beta$ in a given era, and the 
horizon scale will grow even faster ($\propto t$). Thus, in finite time, 
$E$ will no longer be a significant fraction of the horizon 
scale, $\dot{E}$ will decrease, and $E$ will return to being 
constant as in the case of non-expanding space. 
In other words, a large, horizon-scale loop is affected by 
expansion until it falls inside the horizon.

\TABLE{
\begin{tabular}{|c | c | c || c | }
\hline
$\frac{a^2\textbf{x}'^2}{E^2h}$ & $a^2\dot{\textbf{x}}^2$ & 
Internal velocity and length terms & $E$ \\ 
\noalign{\hrule height 2pt}
large & small & small & $\propto a$ \\ \hline
small & large & small & $\propto 1/a$ \\ \hline
small & small & large & $\sim$constant \\ \hline
\end{tabular}
\caption{Approximate behaviour of $E$ when different length- and 
velocity-squared terms are dominant.}\label{expansiontab} }

We see this behaviour in the simple Ansatz (\ref{circleansatz}). 
Evolving numerically from such a circular configuration 
with superhorizon physical radius and zero velocities in all dimensions, 
we see that that $E$ begins by increasing proportionally to the 
scale factor, then oscillates up and down as the circular 
loop oscillates in and out, and approaches a constant value over time 
(Fig.~\ref{Eoft_supermed}). An interesting 
difference with respect to the pure 4D case becomes apparent. 
In 4D, a static, superhorizon loop starts with $\dot E\simeq(\dot a/a)E$ 
and this rate gets gradually reduced over cosmological timescales 
as the loop velocity slowly builds up from 0. Here, there is a 
second scale -- that of the warping -- which is hierarchically smaller than 
the expansion rate, and causes the build up of velocities in the 
compact dimensions over a much 
shorter timescale through the gradient terms in equation (\ref{etadamp2}). 
As a result, the loop energy very quickly turns to a 
significantly slower expansion with $\dot E\sim (1/2)(\dot a/a)E$ 
(see Fig.~\ref{v_eta_supermed}), which can be understood 
from equation (\ref{Edamp2}) with $a^2 \dot{\bf x}^2\to 0$ and 
$h\epsilon^{\frac{4}{3}}\frac{\dot{\eta}^2}{6K^2}\lesssim 
1$\footnote{In fact, the velocity $h^{\frac{1}{2}}
\epsilon^{\frac{2}{3}}\frac{\dot{\eta}}{\sqrt{6}K}$ oscillates in 
the interval $(-1,1)$ and the rate $\dot E$ changes accordingly 
on short timescales. Over larger timescales, the evolution can be 
approximated by taking the root-mean-squared velocity 
which is close to $1/2$.}. Thus, the horizon, 
growing linearly in time, catches up with the physical 
loop size earlier than in 4D.  Note, however, that a 
similar behaviour also occurs in 4D if the initial loop 
has small-scale-structure. In that case, the small-scale 
curvature generates velocity over short timescales and the 
situation is similar to the one we just discussed. 

\FIGURE{
\includegraphics[scale=0.8]{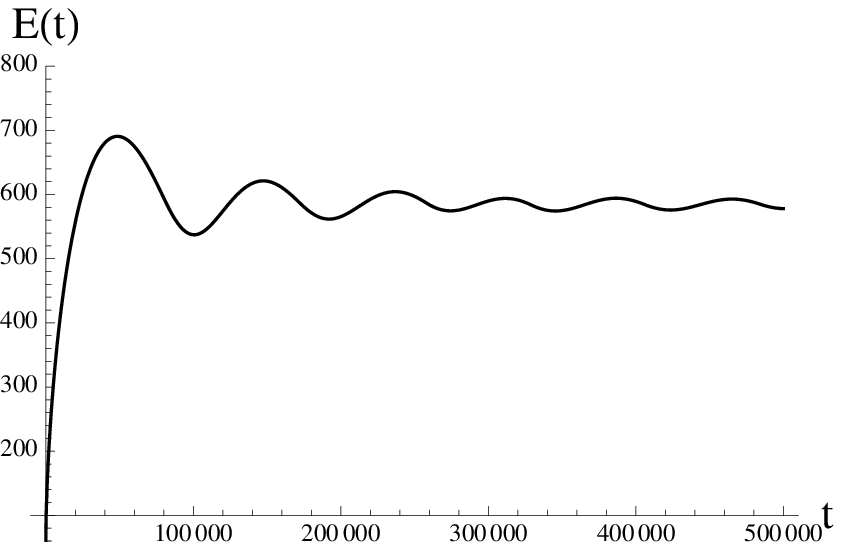}\ \ \ 
\includegraphics[scale=0.8]{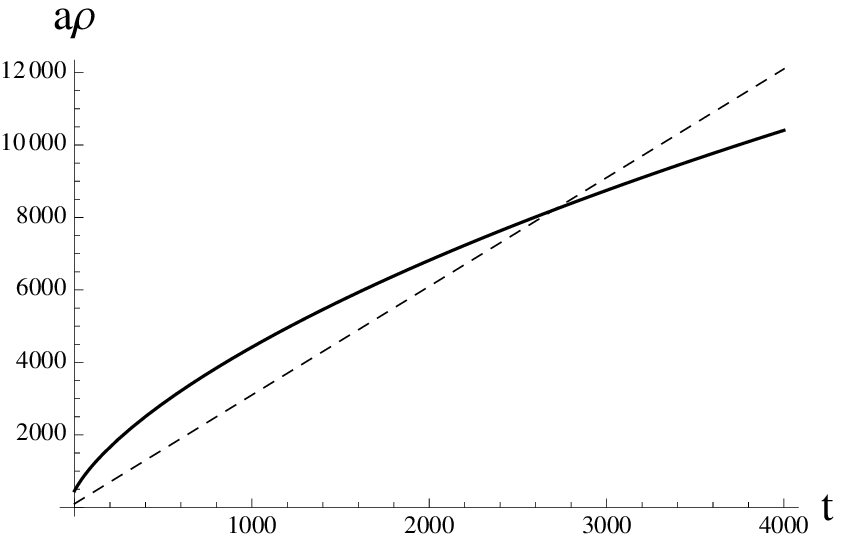}
\caption{Left: Evolution of $E(t)$ for an initially 
static superhorizon loop. $E$ is initially increasing but starts 
oscillating after the loop falls inside the horizon and eventually 
approaches a constant as in flat space. 
Right: Early evolution of the loop's physical radius $a\rho$ (solid line) 
until it falls inside the horizon scale (dashed line). Over larger 
timescales (not shown), $a\rho$ undergoes oscillations with an 
amplitude that gets smaller with respect to the horizon, in 
accordance to the energy plot on the right.}\label{Eoft_supermed}}

\FIGURE{
\includegraphics[scale=0.8]{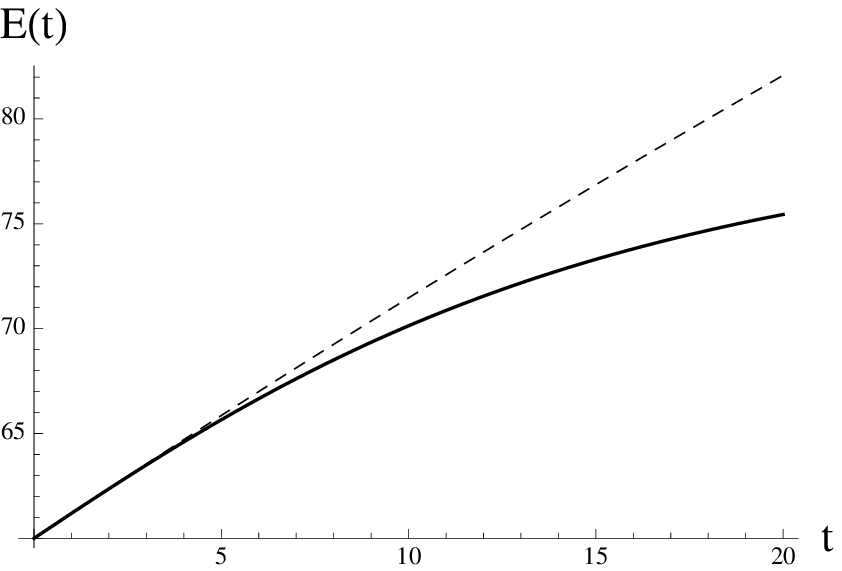}\ \ \ 
\includegraphics[scale=0.8]{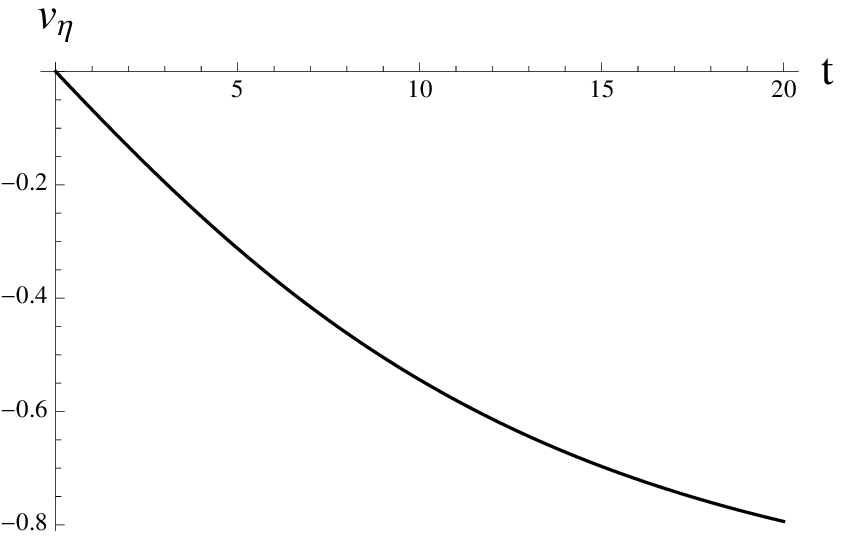} 
\caption{Left: Early time evolution starts as 
$E(t)\propto a(t)$ initially, but quickly drops to a slower rate 
as internal velocities (shown in the right plot) build up. 
The dashed line shows the $E(t)\propto a(t)$ growth, which would 
have been expected if there were no internal velocities.  
Right: Build up of internal velocity $v_\eta\equiv h\epsilon^{\frac{4}{3}}
\frac{\dot{\eta}^2}{6K^2}$ over the same timescale, which, in view of 
equation (\ref{Edamp2}), is responsible for the reduction of 
the rate $\dot E$ in the left plot. Over larger timescales (not shown), 
$v_\eta$ oscillates with $-1\le v_\eta \le 1$ and the 
rate $\dot E$ changes accordingly, so on much larger timescales 
the evolution can be approximated by taking the root-mean-squared 
$\langle v_\eta^2\rangle\simeq 1/2$.}\label{v_eta_supermed}  }

Having looked at the overall behaviour of $E$, we turn our 
attention to the Hubble terms proportional to $\frac{\dot{a}}{a}$ 
that appear in equations (\ref{xdamp2})-(\ref{phidamp2}).
Since these terms scale as $\frac{\dot{a}}{a}\propto \frac{1}{t}$ 
their relative effect  generally becomes less important at later times, 
but they can still dominate during short intervals when 
the string radius crosses zero and $E$ is kinetic energy-dominated.
The term  in the equation for external motion, (\ref{xdamp2}), 
is a damping term, while the terms in the internal equations, 
(\ref{etadamp2}) and (\ref{phidamp2}), can in general have positive or 
negative sign with respect to the corresponding internal velocity.

It is important to note however, that although there 
is a damping term in the equation for \textbf{x}, 
which reduces the size of its oscillations over time, the 
physical variable, $a\textbf{x}$, does not get smaller.
The damping term in equation (\ref{xdamp2}) 
is stronger when the term $\frac{a^2\textbf{x}'^2}{E^2h}$ is larger, 
corresponding to the loop radius being close to a local maximum. 
This implies that the comoving velocity, $\dot{\textbf{x}}$, which 
is on average lowest in such configurations, will be slowed  down 
even more in this region, so 
the string will tend to spend longer in such configurations. Then, looking 
back at the other equations, we see that this corresponds to the region 
where $E$ is increasing and where the internal coordinates, $\eta$ 
and $\phi$, are being damped. This suggests that overall, the energy 
of a loop will slightly increase, and the extra dimensional motion will be 
slightly damped. The overall increase of energy is seen 
in the simple Ansatz, figure~\ref{Eoft_supermed}, and intuitively 
makes sense as the string is initially ``stretched'' by 
the expansion of space. 
The slight damping agrees with \cite{tasos2008}, but again it is not
enough to localise the string.
Then, after the loop falls inside the horizon, $E$ becomes constant, 
and the string behaves exactly as it did in non-expanding 
space, as discussed above.  Thus, we confirm that Hubble damping 
alone cannot localise the strings in the internal dimensions.  
In fact, if $E$ has increased from its starting value, motion in 
both sectors will be larger overall.

Figure~\ref{rhoetavs_supermed} shows the evolution of the 
comoving radius $\rho$ and the radial position $\eta$ in the 
throat (left plot), together with the corresponding velocity 
evolutions (right plot), for the same loop considered above.  
Note that the apparent damping in the motion of the comoving radius 
is compensated by the scale factor growth $a(t)\propto t^\beta$, 
so overall the physical loop radius does not shrink at late times. 
(Here, we neglect gravitational radiation, to be studied in 
the next subsection~\ref{gravradn}.) As we have discussed, 
there is a small damping in the $\eta$ motion, but the amplitude 
increases back again as energy is exchanged between the two sectors. 
\FIGURE{
\includegraphics[scale=0.8]{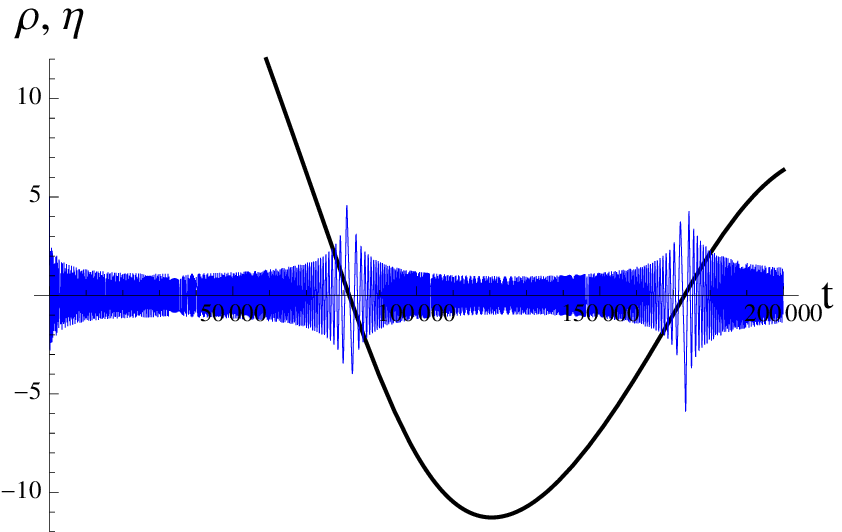}\ \ \ 
\includegraphics[scale=0.8]{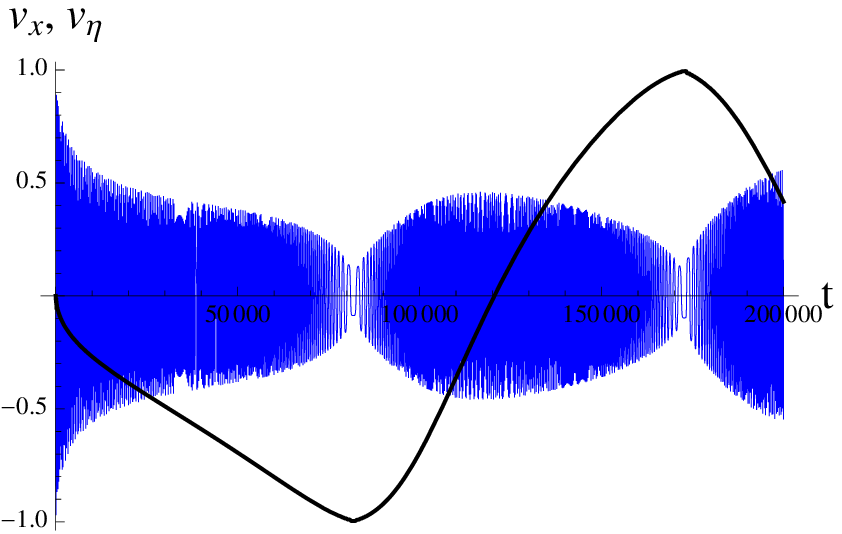}
\caption{Left: Evolution of the comoving loop 
radius $\rho$ (black thick line) and the internal radial 
coordinate $\eta$ (blue thin line) for the initially superhorizon sized
loop discussed above.  The motion of $\eta$ exhibits high-frequency 
oscillations up and down the throat, with an amplitude profile that 
is correlated with the 3D motion of $\rho$.  The amplitude of $\eta$ 
reaches a local maximum when $\rho$ passes from $0$, at which point 
{\bf almost} all kinetic energy is in the 3D motion (cf.~right plot).    
Right: Evolution of the physical velocities $v_x$ (black thick line) 
and $v_\eta$ (thin blue line) corresponding to $\rho$ and $\eta$ 
respectively.  The internal velocity $v_\eta$ oscillates at a 
much higher frequency, determined by the warping, becoming very small
when the 3D velocity becomes relativistic, as the radius, $\rho$, 
passes through $0$ (cf.~left plot).
}\label{rhoetavs_supermed} }

\subsection{Gravitational radiation}\label{gravradn}

Cosmic strings or superstrings lose energy via gravitational 
radiation, and while this is by no means the 
only form of radiation they produce, it is the most 
important for our purposes. For information on other 
forms of radiation, the emission of gamma rays, neutrinos 
and protons is discussed in \cite{partrad}. 
However, this emission is only significant when loops are 
produced at the scale of the string width, and 
recent simulations \cite{loopsize} (but see 
also~\cite{grp1,grp2,grp3}) suggest that while 
there may initially be a high production of small 
loops, a scaling population of large loops at 
approximately $\frac{1}{20}$ of the horizon size becomes 
dominant, for which gravitational radiation is the main form of 
energy loss. 

There has also been work recently on the emission of light 
from cosmic strings \cite{lightrad}, but the 
power radiated by this mechanism is significantly smaller 
than that from gravitational radiation. In the 
context of string theory, cosmic (super)strings can emit many 
sorts of fields (for example \cite{RRrad} suggests that the 
Ramond-Ramond field would be the dominant form of radiation 
for cosmic D-strings) but these will only be produced at extremely 
high energies in the very early universe. Gravitational 
radiation will therefore be the dominant energy-loss 
mechanism for cosmic string loops for the majority 
of their history, so we chose to model this effect. 
Note, however, that other forms of radiation could be 
modelled in a similar way. Finally, note that the internal
excitations on cosmic (super)strings could also be damped 
by their interaction with Kaluza-Klein (KK) gravitons~\cite{Dufaux}. 
However, over cosmological timescales such massive modes 
can be expected to decay, and we do not expect these to
lead to an additional signal in the gravitational wave spectrum. 
While such a process can provide an additional damping mechanism,
this will not qualitatively change any of our conclusions.

The emission of gravitational radiation by cosmic string 
loops can be approximated by a constant rate 
of decrease of the overall energy of the loop, until 
the energy reaches zero and the string has 
disappeared. In flat 4D spacetime, the following formula 
was derived by Vachaspati and Vilenkin \cite{gwave2} for the
rate of energy loss from a loop:
\be \label{quadrupole}
\frac{d\mathcal{E}}{dt} \sim \Gamma G\mu^2 \,,
\ee
where $\mathcal{E}$ is the invariant energy of the loop (defined
in (\ref{relateEE})), $\mu$ the string tension and $G$ is Newton's constant. 
Equation (\ref{quadrupole}) is similar to the quadrupole formula that 
applies to slow-moving sources, with the addition of a numerical 
factor, $\Gamma$, which is usually evaluated to be 
approximately 50, \cite{GammaValue}. 
This factor results from the fact that cosmic string loops are 
fast-moving, and, in particular, part of the contribution to $\Gamma$ 
comes from cusps. As described 
in the introduction, these are events at which a 
point on the string instantaneously approaches the speed of light, 
and are generic on smooth loops in 4D. As discussed in \cite{CCGGZ,CCGGZ2}, 
allowing the string to move in extra dimensions greatly reduces 
the occurrence of cusps and may therefore affect the value of $\Gamma$.
This would likely reduce the effective $\Gamma$, so taking $\Gamma\simeq 50$, 
could well be over-estimating the damping due to gravitational radiation.

The general behaviour, however, is independent of the numerical factor. 
The magnitude of the energy affects both how far away from the tip of 
the throat the string tends to move, and the maximum length of 
the loop. From (\ref{relateEE}) and (\ref{Efull}), the energy is:
\be \label{gravineq}
\mathcal{E} = \mu\int d\sigma \sqrt{\frac{\textbf{x}'^2
+h\epsilon^{\frac{4}{3}}\left(\frac{\eta'^2}
{6K^2}+B\phi'^2\right)}
{ h\left(1-\dot{\textbf{x}}^2-h\epsilon^{\frac{4}{3}}
\left(\frac{\dot{\eta}^2}{6K^2}+B\dot{\phi}^2\right)
\right) } }
\geq \mu\int d\sigma \sqrt{\frac{\textbf{x}'^2
+h\epsilon^{\frac{4}{3}}\left(\frac{\eta'^2}{6K^2}+B
\phi'^2\right)} { h(0) } } \,.
\ee
The far right hand side of this equation is proportional to 
the total length of the string loop, from which 
one can see that as $\mathcal{E}$ decreases, the constraint 
on the maximum length of the loop becomes tighter. 

There is no explicit constraint on $\eta$ (distance from 
the tip of the throat), since if the length of the 
string is zero at any point, $\eta$ can be arbitrarily large. 
However, the length of a loop will generally be 
non-zero, which gives a constraint on $\eta$ and confines the 
loop within a certain distance from the tip of 
the throat. From (\ref{gravineq}) we have:
\be \label{gravineq2}
\mathcal{E} \geq \mu\int d\sigma \sqrt{\frac{\textbf{x}'^2
+h\epsilon^{\frac{4}{3}}\left(\frac{\eta'^2}
{6K^2}+B\phi'^2\right)} { h(\eta) } }
\ee
Again, the terms in the numerator give the length of the loop. 
Since $h$ is a decreasing function of $\eta$, this equation 
implies that the restriction on $\eta$ becomes tighter as $\mathcal{E}$ 
gets smaller. Using an analytic approximation
for $h$ we can calculate the rate at which the restriction tightens. 
As $\mathcal{E}$ tends towards zero (as the loop loses energy), we 
find that the restriction on $\eta$ does not tighten as quickly 
as the restriction on the length of the loop. 
In fact, when $\mathcal{E}$ reaches zero, the total length of 
the loop must be zero, whereas there is no 
requirement for the string to be at the bottom of the 
throat ($\eta=0$). From this we argue that, in 
general, the length of the loop disappears before the internal 
motion, so it will not be localised at the tip 
of the throat. 
\FIGURE{
\includegraphics[scale=0.75]{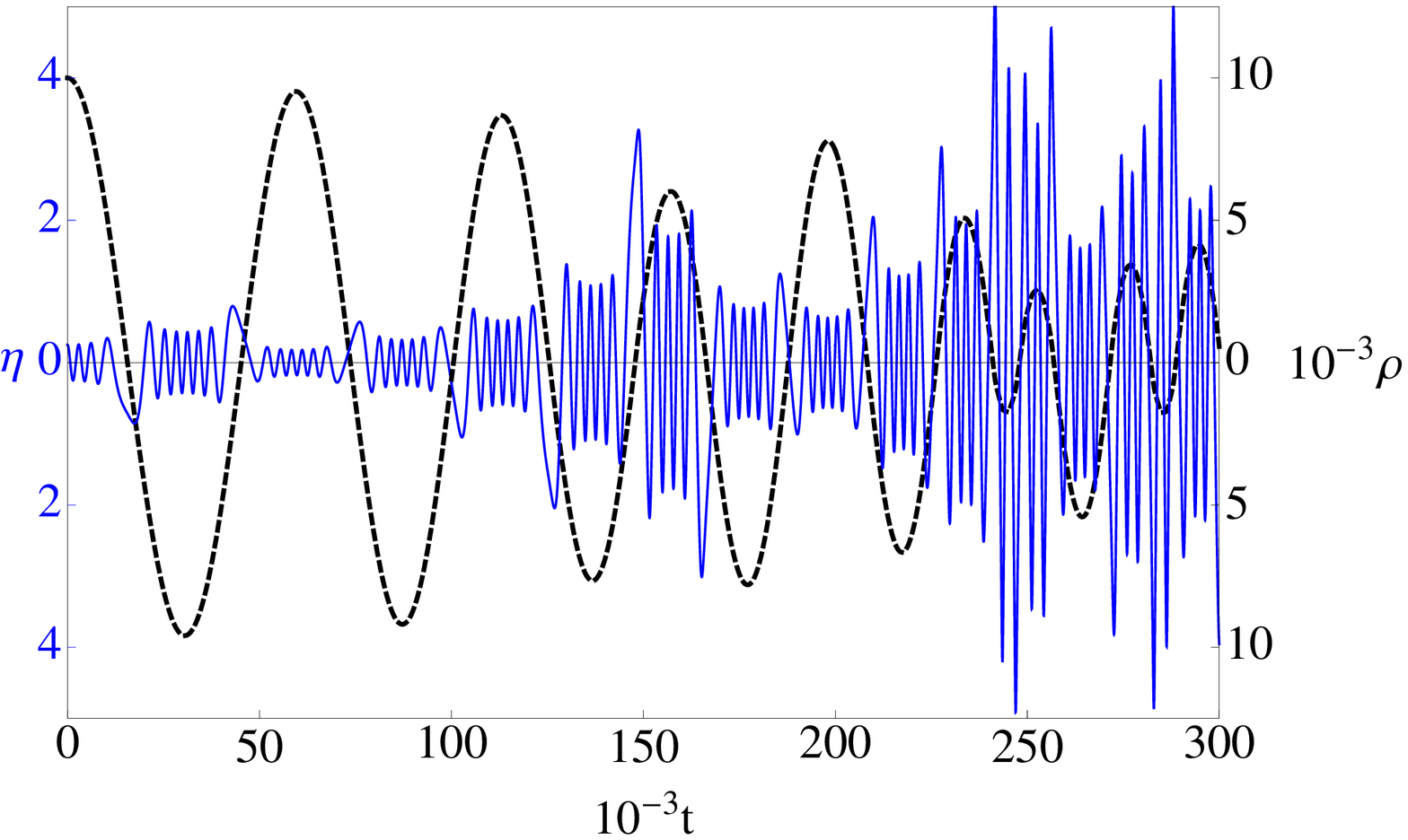} 
\caption{Including the gravitational radiation damping into the circular
loops of section \ref{ansatzsec}. The loop shown has no angular momentum, and
corresponds to the damped version of the first plot in figure \ref{fig:withJ}.
A rather large value of damping was chosen, corresponding
to a halving of $E_0$ by the end of the integration.
}
\label{fig:gravplot} }

Figure \ref{fig:gravplot} shows the effect of the gravitational
radiation damping on the simple loop Ansatz (\ref{circleansatz}).
A rather large value of damping, equivalent to setting $G\mu\simeq 10^{-7}$
in (\ref{quadrupole}), was chosen
to highlight the effect, and demonstrates nicely how the external
loop radius, $\rho$, reduces overall more quickly than the internal 
oscillations. The motion towards the end of the time period becomes
more and more noisy as the loop samples more the sharp valley in the
effective potential at $\rho=0$. 
We conclude that the emission of gravitational radiation will 
neither cause the string to be pulled to the tip of the throat 
nor the motion to become effectively 4D.

\section{Discussion} \label{discussion}

In this paper, we have explored explicitly the motion of a 
cosmic (super)string on a compactification with warped internal
dimensions, using the Klebanov-Strassler throat as a test
geometry. The results show that there is no classical geometrical or
dynamical mechanism which preferentially damps the motion in 
the internal dimensions. On the contrary, we observe a general
tendency for motion in the internal dimensions to either become
or persist in being appreciable under a wide range of situations.
We now discuss the ways in which motion differs from the case 
where the extra dimensions are flat, and how this impacts the
gravitational wave signal.

Taking a flat metric allows a third gauge choice to be made 
along with the transverse temporal gauge, so we take the following:
$$
\tau = t,\ \dot{x}^Mx'_M = 0,\ \dot{x}^2 + x'^2 = 0 \,.
$$
The equations of motion for the string then become very simple. 
The trajectory in the spatial directions, 
which are all flat, is simply given by a wave equation:
\be \label{flateom}
\ddot{\textbf{x}} = \textbf{x}'' \,.
\ee
This has an analytic solution that is simply a sum of left 
and right-moving waves:
$$
\textbf{x}(\sigma,t) = \frac{1}{2}\left(\textbf{a}(\sigma-t)
+\textbf{b}(\sigma+t)\right) \,.
$$
This implies, since a closed loop must be periodic in $\sigma$, 
that it must also be periodic in $t$. The functions \textbf{a} 
and \textbf{b} can take any form subject to the following 
constraints (which follow from the gauge conditions):
$$
\textbf{a}'^2 = \textbf{b}'^2 = 1 \,.
$$
Here, prime represents differentiation by their respective arguments. 
Thus, with $n$ spatial directions, these functions are constrained 
to lie on a unit $(n-1)$-sphere. This gives some interdependence 
between the different directions. Essentially it means that 
if $\textbf{a}'$ or $\textbf{b}'$ is large in one 
direction, it must be small in another. This is directly related 
to the conservation of energy, given by the 
condition: $ \textbf{x}'^2 + \dot{\textbf{x}}^2 = 1 $, which 
implies that if length or velocity terms are large 
in one direction, they must be small in another.

In the case of warped spacetime, loop trajectories are no longer 
periodic (as we see clearly on figure \ref{fig:withJ}, for example). 
This is essentially due to the presence of two different coupled
characteristic frequencies of oscillation: the characteristic 
frequency of the loop based on its size, and the frequency of 
oscillation up and down the throat, based on the scale of the throat. 
These two different frequencies interfere, resulting in non-periodic motion. 
This effect is particularly noticeable if the frequencies are 
comparable, which depends on the scale of the loop and the 
size of $\epsilon$ (which gives the scale of the throat).

In a similar way to the flat space case described above, the 
length and velocity terms are still constrained by the conservation 
of energy so that if they are large in a particular direction they must be 
small in another, which is expressed by the following equation:
$$
\dot{\textbf{x}}^2 + h\epsilon^{\frac{4}{3}}
\left(\frac{\dot{\eta}^2}{6K^2}+B\dot{\phi}^2\right)
+ \frac{\textbf{x}'^2}{E^2h} + \frac{\epsilon^{\frac{4}{3}}}{E^2}
\left(\frac{\eta'^2}{6K^2}+B\phi'^2\right) = 1
$$
However, there is an additional interaction between the 
different directions that is not present in flat space. 
We saw in section \ref{generalloops} that motion in all 
directions is affected by the radial distance, $\eta$, 
of the string from the tip of the warped throat. 
We saw that if a string is close to the tip 
of the throat, its total length and velocity are larger, 
and they are smaller if the string is far from the tip. When
the frequency of internal oscillations is much higher than those
in the external part of the loop, which is the case for a 
realistic cosmological
scenario, this results in more of an averaged effect (see for example
figure \ref{fig:withJ}), but nontheless an important one.

The fact that some interaction between the internal radial 
distance and the other directions exists is 
apparent in the equations of motion, since the equation of 
motion (\ref{etadamp}) for the internal radial distance, $\eta$, 
contains terms depending on all the other 
directions, while the other equations of 
motion, (\ref{phidamp}) and (\ref{xdamp}), depend on $\eta$. 
In (\ref{flateom}), we see that no such 
dependence appears in flat spacetime.

This all means that strings in warped spacetime have more 
interesting and complex motion than those 
in flat spacetime, and we have seen a few examples of 
their intruiging dynamics in this paper (see for 
example figures \ref{pdetraj1} and \ref{pdetraj2}).
The main conclusion is that the warped geometry will
also contain appreciable internal motion which will then
impact on the physics of the 4D cosmological network.

In \cite{CCGGZ,CCGGZ2}, it was shown that if there was a portion of kinetic
energy in the internal dimensions, then the string would slow
down in our external noncompact dimensions. This then feeds into
a reduced gravitational wave signal as the energy-momentum of the string,
\be
T^{\mu\nu} = \mu \int d^2 \sigma ({\dot X}^\mu {\dot X}^\nu -
X'^\mu X'^\nu ) \; \delta ^{(4)} (x^\mu - X^\mu ( \tau, \sigma ) \,,
\ee
is sharply peaked in fourier space around the wave vectors of
the left and right moving modes. It is only when these wave vectors
either coincide (as in a cusp) or one has a discontinuity (as in a kink)
that the contribution to the linearized gravitational wave
is enhanced. If the wave vectors cease to be null, then there is 
a corresponding lowering of the overall power in a gravitational
wave burst, as well as a narrowing of the region in which it is beamed.
The near cusp event parameter, $\Delta$, measuring how close to
relativistic the wave velocities of the string are, is the parameter
which governs the reduction in the signal. 

In this paper we explored the values of $\Delta$ for a family of simple
loops, where we could identify where the string was moving at its most
relativistically. This gave sample data for $\Delta$ which ranged
(roughly) from $10^{-2}-10^{-3}$. In \cite{CCGGZ2}, the
overall effect on the gravitational wave signal for cusps was 
obtained by marginalising over the $\Delta$ parameter up to a maximum
value $\Delta_0$. The results here indicate that $\Delta_0 \sim 10^{-3}$ 
might be a conservative but sensible value to take, in which case 
(referring to figure 6 of \cite{CCGGZ2}), we see that the expected 
detection rate of a cusp event is likely to be around 1 per century!

\acknowledgments

We would like to thank Louis Leblond for helpful discussions.
AA is supported by a Marie Curie IEF Fellowship at the the 
University of Nottingham.
SC is supported by an EPSRC studentship.
RG is supported in part by STFC (Consolidated Grant ST/J000426/1),
in part by the Wolfson Foundation and Royal Society, and in part
by Perimeter Institute for Theoretical Physics.
Research at Perimeter Institute is supported by the Government of
Canada through Industry Canada and by the Province of Ontario through the
Ministry of Economic Development and Innovation.
AA and SC would like to acknowledge the support of the CTC at
Cambridge, and also SC the hospitality of the Perimeter Institute 
during the course of this project.

\end{document}